\documentclass[fleqn,usenatbib]{mnras}

\usepackage{newtxtext,newtxmath}
\usepackage[T1]{fontenc}
\usepackage{ae,aecompl}
\usepackage{graphicx}
\usepackage{amsmath}
\newcommand{\angstrom}{\textup{\angstrom}}
\usepackage{amssymb}
\usepackage{multirow}

\newcommand\lsim{\mathrel{\rlap{\lower4pt\hbox{\hskip1pt$\sim$}}
        \raise1pt\hbox{$<$}}}
\newcommand\gsim{\mathrel{\rlap{\lower4pt\hbox{\hskip1pt$\sim$}}
        \raise1pt\hbox{$>$}}}

\defcitealias{Dorazio2015Nature}{DHS15}

\title[Multi-wavelength test of Doppler boost for PG1302-102]{Testing the relativistic Doppler boost hypothesis for the binary candidate quasar PG1302-102 with multi-band \\{\it Swift} data}

\author[C.~Xin et al]{Chengcheng~Xin,$^{1}$
Maria~Charisi,$^{2}$\thanks{E-mail: mcharisi@caltech.edu}
Zolt{\'{a}}n~Haiman,$^{1}$ David Schiminovich,$^{1}$ \newauthor Matthew~J.~Graham,$^{2}$ Daniel~Stern,$^{3}$ Daniel J. D'Orazio$^{4}$ 
\\
$^{1}$Department of Astronomy, Columbia University, New York, NY, 10027\\
$^{2}$Division of Physics, Mathematics and Astronomy, California Institute of Technology, Pasadena, CA, 91125\\
$^{3}$Jet Propulsion Laboratory, California Institute of Technology, Pasadena, CA, 91109\\
$^{4}$Astronomy Department, Harvard University, Cambridge, MA 02138}

\date{Accepted XXX. Received YYY; in original form ZZZ}

\pubyear{2019}

\begin{document}
\label{firstpage}
\pagerange{\pageref{firstpage}--\pageref{lastpage}}
\maketitle

\begin{abstract}
The bright quasar PG1302-102 has been identified as a candidate supermassive black hole binary from its near-sinusoidal optical variability. While the significance of its optical periodicity has been debated due to the stochastic variability of quasars, its multi-wavelength variability in the ultraviolet (UV) and optical bands is consistent with relativistic Doppler boost caused by the orbital motion in a binary. However, this conclusion was based previously on sparse UV data which were not taken simultaneously with the optical data. Here we report simultaneous follow-up observations of PG1302-102 with the Ultraviolet Optical Telescope on the {\it Neil Gehrels Swift Observatory} in six optical + UV bands. 
The additional nine {\it Swift} observations produce light curves roughly consistent with the trend under
the Doppler boost hypothesis, which predicts that UV variability should track the optical, but with a $\sim2.2$ times higher amplitude.
We perform a statistical analysis to quantitatively test this hypothesis. We find that the data are consistent with the Doppler boost hypothesis when we compare the the amplitudes in optical $B$-band and UV light curves.
However, the ratio of UV to $V$-band variability is larger than expected and is consistent with the Doppler model, only if either the UV/optical spectral slopes vary, the stochastic variability makes a large contribution in the UV, or the sparse new optical data underestimate the true optical variability. We have evidence for the latter from comparison with the optical light curve from ASAS-SN. Additionally, the simultaneous analysis of all four bands strongly disfavors the Doppler boost model whenever {\it Swift} $V$-band is involved. Additional, simultaneous optical + UV observations tracing out another cycle of the 5.2-year proposed periodicity should lead to a definitive conclusion.
\end{abstract}


\section{Introduction} 
\label{sec:intro}
It is well established that all massive galaxies host supermassive black holes (SMBHs), with masses $10^6 - 10^{10} M_\odot$, in their nuclei  \citep{Kormendy2013}. 
According to cosmological models of structure formation, galaxies merge frequently to form more massive galaxies (e.g., \citealt{Haehnelt2002}). It follows that compact SMBH binaries (SMBHBs) should be common in galactic nuclei~\citep{Begelman1980}. As a by-product of galaxy mergers, SMBHBs are important for understanding galaxy evolution. They are also important because at small (milli-parsec) separations, they become strong sources of low-frequency gravitational waves (GWs), and are the prime targets for experiments like Pulsar Timing Arrays (PTAs; e.g., \citealt{Burke_Spolaor2019}) and the {\it Laser Interferometer Space Antenna (LISA)}.\footnote{See http://lisamission.org}

Despite their expected ubiquity,  observational evidence, especially for compact sub-parsec SMBHBs, remains sparse \citep{DeRosa2020}.
Dual AGN at kpc separations have been repeatedly resolved in X-rays, optical and infrared  \citep{Komossa2003,Comerford2011}, but as the SMBHs move to smaller separations, they can only be resolved in radio bands, with Very Long Baseline Interferometry (VLBI; e.g., \citealt{Rodriguez2006}).
At sub-parsec separations, they are practically below the resolution limits of even VLBI (although see \citealt{Dorazio_2017_VLBI}). 
Therefore, the presence of a binary needs to be inferred indirectly from its effect on the surrounding matter.\footnote{We also note that, in the (not-too-distant) future, compact binaries will be directly ``observable" in GWs with PTAs and {\it LISA}.}

One proposed method to identify SMBHBs is to search for periodic variability in quasars. The intuitive expectation that the orbit of the SMBHB will periodically perturb the nearby gas has been confirmed in multiple hydrodynamical simulations. 
Overall, the emerging picture is that the binary evacuates a central cavity in the disc, while gaseous streams enter the cavity periodically and efficiently accrete onto the SMBHs \citep{AL96,MM08,Cuadra+2009,Roedig+11,Nixon+2011,Roedig+2012,DOrazio2013,Gold+2014}.
This likely results in bright quasar-like luminosity (possibly, repeating bursts), which is periodically modulated at roughly the orbital period of the binary, with the structure of the periodogram depending strongly on the mass ratio (see, e.g.,  \citealt{Farris+2014,ShiKrolik2015,Dorazio+2016}).

Additionally, some of the incoming gas becomes bound to the SMBHs, creating mini-discs around each SMBH \citep{Ryan2017,Tang2017}. In compact binaries, the SMBHs move at relativistic speeds, and thus the emission from the mini-discs, and in particular, the emission of the secondary mini-disc (which is expected to be brighter and has a higher orbital velocity) is Doppler boosted.
Since the thermal disc emission is radially stratified, with higher-energy emission arising from smaller radii, the expectation is that at higher frequencies, this Doppler effect is increasingly important, and becomes dominant above frequencies corresponding to thermal emission from the outer edges of the mini-discs (roughly coinciding with the $V$ band in the case of PG1302-102; see \citealt{Dorazio2015Nature}).
For typical spectral slopes $\alpha_{\nu}\equiv d\ln F_\nu /d\ln\nu<3$ 
the binary will appear brighter/dimmer, when the secondary SMBH is approaching/receding from the observer, even if the rest-frame luminosity is constant.  To first order in orbital velocity and for a power-law spectrum, the observed flux is modulated as
\begin{equation}
\label{eq:deltaFnu}
   \frac{\Delta F_{\nu}}{F_{\nu}}=(3-\alpha_{\nu})\frac{v}{c}\cos\phi\sin i,
\end{equation}
where $v$ is the orbital velocity of the more luminous SMBH (with the other BH assumed to be much dimmer), $i$ is the inclination of the orbit with respect to the line-of-sight and $\phi$ is the orbital phase. For unequal-mass binaries that are not too far from edge-on, the Doppler boost may dominate the variability, producing a smooth quasi-sinusoidal light curve.

Systematic searches for quasars with periodic variability in time-domain surveys, e.g., the Catalina Real-Time Transient Survey (CRTS), the Palomar Transient Factory (PTF), and the Panoramic Survey Telescope and Rapid Response System (Pan-STARRS) have identified $\sim$150 binary candidates \citep{Graham2015b,Charisi2016,Liu2019}.\footnote{Additional candidates have also been identified individually \citep{Zheng2016,Bon2016,Li2019,Dorn2017}, although the statistical significance of the latter was brought into question by \citet{Barth2018}.}
However, the stochastic variability of quasars can introduce spurious detections. This is further aggravated by our incomplete understanding of the precise form of intrinsic quasar variability \citep{Vaughan2016} coupled with the relatively short baselines, in which only a few cycles can be observed. 
Indeed, several recent studies have found inconsistencies with the widely used Damped Random Walk (DRW) models, and favored other descriptions of stochastic quasar variability (e.g. \citealt{Mushotzky+2011,Caplar2017,Smith+2018}). Throughout our analysis, we here nevertheless follow  \citet{Graham2015b} and \citet{Charisi2016}, who explicitly included stochastic noise in their statistical analysis by assuming that quasar variability is described by a DRW model.

\citet{Sesana2018} demonstrated that the samples of quasars with periodic variability likely contain many false positives. They found that the GW background inferred from this population of binary candidates is in tension with the PTA upper limits. On the other hand, theoretical models predict that at least a few closely separated SMBHBs should be detectable in the current time-domain surveys \citep{Haiman2009,Kelley2019}. It is thus crucial to select the genuine binaries among the candidates by identifying additional binary signatures, such as multiple components of periodic variability \citep{Charisi2015,Dorazio2015}, self-lensing flares \citep{Dorazio2018}, or the wavelength dependence of the Doppler modulation \citep{Dorazio2015Nature, Charisi2018}.

Among the identified candidates, a prominent source from the CRTS sample is quasar PG1302-102  \citep{Graham2015a}. It is a bright quasar at redshift $z=0.27$ with a BH mass of $\sim10^{9} M_\odot$. It exhibits quasi-sinusoidal variability with a period of $\sim$5.2\,yr and an amplitude of $\sim$0.14\,mag in $V$-band. The significance of the periodicity has been a topic of controversy; \citet{Vaughan2016} with a Bayesian analysis showed that the DRW model is preferred to a sinusoid, whereas \citet{Dorazio2015Nature} with a similar approach reached the opposite conclusion.\footnote{The different conclusions are possibly due to the dramatically different best-fitting $\tau$ parameters for the DRW model.} \citet{Charisi2015} also found the periodogram peak to be significant, but only considering it as a stand-alone detection (i.e. trial factors, to account for the fact that PG1302-102 was chosen from a large sample, were not included).
Recently, \citet{Liu2018} added data from the All-Sky Automated Survey for Supernovae (ASAS-SN) and found that a sinusoidal+DRW model is preferred to a pure DRW model, but the significance of the periodicity decreased.\footnote{The light curve from ASAS-SN has inferior photometric quality compared to CRTS. Also, the binning of the light curve may significantly affect the statistical analysis. \citet{Liu2018} chose wide bins of 150\,d, longer than the typical DRW time-scale.} Undoubtedly, long-term monitoring will determine whether the periodicity of PG1302-102 is persistent.

Beyond the simple periodicity, \citet{Dorazio2015Nature} [{\it hereafter} \citetalias{Dorazio2015Nature}] suggested that the multi-wavelength variability of PG1302-102 is consistent with relativistic Doppler boost, serving as an additional indication for its binary nature.\footnote{Further signatures for the binary nature of PG1302-102 have been suggested; for instance, its variability in the mid-infrared is quasi-sinusoidal   \citep{Jun2015}, and the angle of its radio jet varies roughly at the proposed optical 5.2-yr period,  \citep{Qian2018}.} More specifically, in the Doppler boost scenario described above, there is a robust multi-wavelength prediction: if the UV luminosity also arises in the mini-discs, the optical and UV light curves should vary in tandem. The variability amplitudes $A_{\rm UV}$, $A_{\rm opt}$ depend on the respective spectral indices $\alpha_{\rm UV}$, $\alpha_{\rm opt}$ (eq. \ref{eq:deltaFnu}), which means that the relative amplitudes in the two bands are
\begin{equation}
    \frac{A_{\rm UV}}{A_{\rm opt}}=\frac{3-\alpha_{\rm UV}}{3-\alpha_{\rm opt}}. \label{eq:1}
\end{equation}
The model prediction was tested with UV spectra and photometry from the {\it Hubble Space Telescope (HST)} and the {\it GALaxy Evolution EXplorer (GALEX)}. However, the UV data were quite sparse in \citetalias{Dorazio2015Nature}. Subsequently, \citet{Charisi2018} demonstrated in a sample of non-periodic quasars that, with the currently available sparse UV data, the multi-wavelength Doppler signature can be confused with wavelength-dependent variability of quasars. The probability that the multi-wavelength Doppler boost signature arises by chance increases as the quality of the UV data decreases (e.g., from 20\% in the near-UV sample to $\sim$40\% in the far-UV sample--see also Figure~2 and 3 in \citealt{Charisi2018}).  These probabilities reflect the limited quality of the data in the control samples, and represent only upper limits on how frequently quasars mimic the Doppler brightness+colour variations by chance.

Motivated by this, we obtained multi-wavelength follow-up data with the Ultraviolet/Optical Telescope (UVOT), on-board the {\it Neil Gehrels Swift Observatory}.
In this paper, we report the new observations and further test the Doppler boost hypothesis by examining whether UV variability tracks that of the optical, but with a larger amplitude.  We assume that the variability of  PG1302-102 consists of a sinusoidal modulation caused by the relativistic Doppler boost with UV and optical amplitudes defined by the spectral slopes in each band, as well as stochastic DRW variability with amplitudes that may differ in each band, and photometric noise. We confront this model with new data points we acquired in two optical and two UV bands at nine distinct epochs. With simulations we assess the probability that the data are consistent with the Doppler boost model by comparing the UV/optical variability ratios.

The rest of this paper is organised as follows.  In \S~\ref{sec:data_analysis}, we describe the new {\it Swift} data, and the details of our statistical analysis.  In \S~\ref{sec:results}, we present the results of our statistical tests, which are discussed further in \S~\ref{sec:discussion}.  We summarise our main conclusions in \S~\ref{sec:summary}.

\label{sec:Data}
\begin{figure*}
 \centering
 \includegraphics[width=0.8\textwidth]{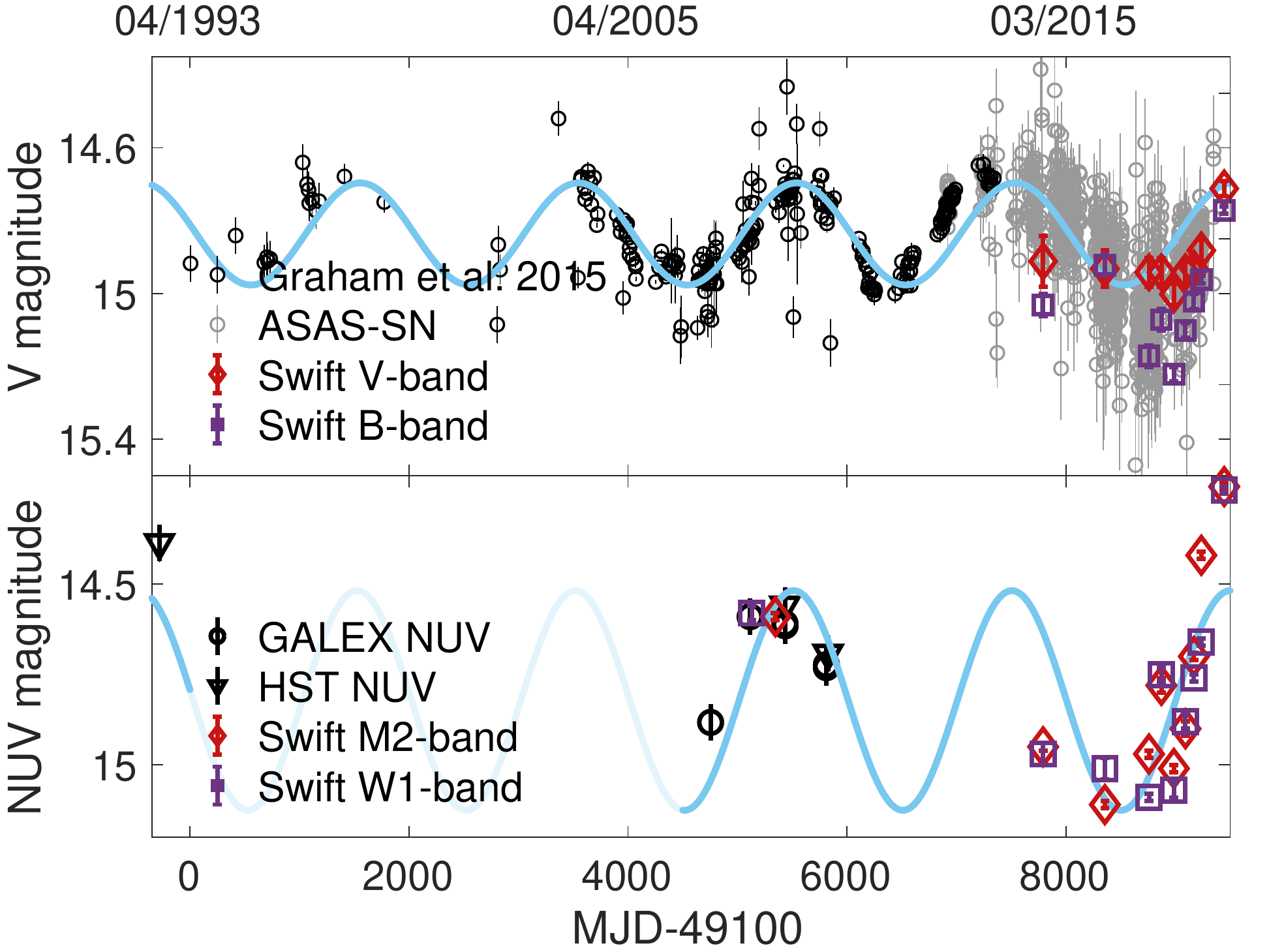}
 \caption{Top Panel: Optical light curve of PG1302-102 with data from \citet{Graham2015a} in black (CRTS+LINEAR and other archival observations), data from ASAS-SN in grey, and purple squares/red diamonds for {\it Swift} $B$/$V$-band observations. Bottom panel: Near-UV light curve, with black circles and triangles for {\it GALEX} and {\it HST} observations from \citetalias{Dorazio2015Nature}, purple squares and red diamonds for {\it Swift} $W1$ and $M2$-band observations, respectively. The sinusoidal Doppler boost model from \citetalias{Dorazio2015Nature} is also shown in light blue.}
 \label{fig:1}
\end{figure*}

\section{Data Analysis}
\label{sec:data_analysis}

\subsection{Data}

We obtained multi-wavelength observations of PG1302-102 with the UVOT on {\it Swift}, initially as a Target of Opportunity, and subsequently, through two approved Guest Observer programs in Cycles 13 and 14 (PI: Z.~Haiman). We extracted the {\it Swift} light curves using the {\it On-line XRT \& UVOT data analysis} pipeline.\footnote{\url{http://www.ssdc.asi.it/mmia/index.php?mission=Swiftmastr}} 
Our observations cover all six filters of UVOT ($B$ and $V$ in optical, $W2$, $M2$, $W1$ and  $U$ in UV). We include one additional archival observation, which also covers all six bands. The photometric measurements of the {\it Swift} optical and UV bands are reported in Table~\ref{Table:swift_data} in Appendix~\ref{sec:appendix-data}.

In Figure~\ref{fig:1}, we present the optical and near-UV light curves of PG1302-102 from our monitoring campaign with {\it Swift}/UVOT, along with archival data from other surveys. More specifically, in the top panel, we show the optical light curve from \citet{Graham2015a} in black, the ASAS-SN light curve,\footnote{We extracted the ASAS-SN light curve from the on-line database \emph{Sky Patrol} \citep{Shappee2014,Kochanek2017}} which was analysed in \citet{Liu2018} in grey. The {\it Swift} observations are superimposed with purple squares for $B$-band and red diamonds for $V$-band. The ASAS-SN and {\it Swift} $V$-band light curves are calibrated in the same photometric system and are directly comparable (see \S~\ref{sec:pipeline}), whereas for the light curve from \citet{Graham2015a}, a constant shift is necessary. We calculated this offset from the difference of the mean magnitudes in the overlapping time interval.

In the bottom panel, we present the near-UV data 
from \citetalias{Dorazio2015Nature} (black circles and triangles for {\it GALEX} and {\it HST} observations, respectively) and the {\it Swift} data points with red diamonds for $M2$-band and purple squares for $W1$-band. Similarly to the optical, we apply a constant offset based on the two {\it Swift} data points that are almost coincident in time with the {\it GALEX}/{\it HST} observations (at MJD$\sim$54,500).
We also show the sinusoidal model for relativistic Doppler boost using the best-fitting orbital parameters from \citetalias{Dorazio2015Nature}.

We note that the {\it Swift} $V$ and $M2$ bands have very similar wavelength coverage to the optical and near-UV bands examined in \citetalias{Dorazio2015Nature} (see Figure~\ref{Fig:UV_Spec} and \ref{Fig:Opt_Spec} below). This allows us to directly compare the new observations with the archival data. It also justifies the choice of a constant offset for the calibration of the different pieces of the time series, since the colour-dependent variability of quasars should have minimal impact in the almost identical filters.

As can be seen from Figure~\ref{fig:1}, the {\it Swift} data cover a total of nine epochs, separated by approximately 3-4 months (over the past two years of our monitoring campaign) and span a baseline of $\sim 1770$ days. 
A key characteristic of our observations is that the data in the distinct filters were taken nearly-simultaneously. This is crucial, because quasars show short-term fluctuations. In previous work, this presented a limitation, since the UV data had to be compared with the extrapolated optical variability. For this reason, we exclude from the analysis a few archival observations that cover only one band. The simultaneous coverage allows more flexibility to test the Doppler hypothesis, beyond the simplest assumption of sinusoidal variability, which corresponds to constant luminosity in the mini-discs.
From hydrodynamic simulations, we expect fluctuations in the accretion rate on shorter time-scale than the orbital period, and thus the intrinsic luminosity of the mini-discs likely may deviate from constant \citep{Farris+2014}.

\subsection{Analysis} \label{sec:Analysis}
The relativistic Doppler boost model predicts the modulation of the observed flux (eq.~\ref{eq:deltaFnu}). 
In the limit of small fluctuations (a reasonable approximation for PG1302-102's O(10\%) variability), to first order the additive magnitude variation is $\Delta m \equiv m-m_0 = \Delta F/F_0$.
In other words, the fractional change in flux and the change in the apparent magnitude are equivalent. We adopt this approximation for the rest of the paper.

We want to test whether the optical/UV variability of PG1302-102 follows the multi-wavelength prediction of the relativistic Doppler model, i.e. the optical and UV light curves vary simultaneously, with amplitudes according to eq. (\ref{eq:1}). 
Our null hypothesis is that the observations are consistent with Doppler boost plus DRW variability. The former reflects the emission of the binary orbiting with relativistic speed, whereas the latter represents additional variability from accretion processes in the quasar.

We quantify the relative change in magnitude between two bands (e.g., $V$ and $M2$), by taking the ratio of the magnitude difference between two observations, i and j (at times $t_i$ and $t_j$, respectively).
\begin{equation}
\label{eq:ratio}
R_{\rm ij} = \frac{\Delta V_{\rm ij}}{\Delta M2_{\rm ij}}=\frac{V(t_i)-V(t_j)}{M2(t_i)-M2(t_j)} 
\end{equation}
We consider the differences between all possible combinations of data points; with 9 distinct observations from {\it Swift}, there are 36 total combinations (without repetition). 
As eq. (\ref{eq:ratio}) implies, we first examine the $V$ and $M2$ bands for comparison with \citetalias{Dorazio2015Nature} (see \S~\ref{sec:Data}), but subsequently we consider multiple combinations of bands. 
Also, in this analysis, we did not include the archival observations from \citetalias{Dorazio2015Nature}, because the optical and UV data were not taken simultaneously.

Introducing the ratio $R_{\rm ij}$ as a metric for the relative change in variability is advantageous for the following reasons.
First, unlike a least-squares fit (or any other similar model fit), our approach does not explicitly make an assumption about the shape of the periodicity. As a result, deviations from a sinusoid, e.g. due to an eccentric binary orbit, fluctuations in the luminosity of the mini-discs, or significant gas motions contributing to the Doppler effect on top of the binary's orbital motion~\citep{Tang+2018}  are automatically incorporated. Additionally, fitting a model with multiple parameters when we only have 9 observations can be problematic (e.g., susceptible to outliers). Another significant advantage is that we do not need to subtract the uncertain mean magnitude in each band. This is especially important in our case, because in addition to having a limited number of observations, we preferentially sample the dim phase of the periodicity. We note that even though the baseline of the {\it Swift} observations is $\sim$1770\,d, close to the detected period of PG1302-102, our dedicated monitoring in Cycles 13 and 14 covers only two years.

In the most idealised case of Doppler boost emission (i.e. constant luminosity of the mini-discs, without any extra intrinsic variability (e.g. DRW) from the quasar, and perfect observations without photometric errors), the ratio $R_{\rm ij}$ would be exactly constant and equal to $(3-\alpha_V)/(3-\alpha_{M2})$, which for the $V$ and $M2$ bands is 1/2.17 \citepalias{Dorazio2015Nature}. However, both the photometric errors and the DRW variability add scatter around the expected value, producing a distribution of $R_{\rm ij}$ values.

In order to assess whether the observed distribution is consistent with the null hypothesis,
we simulate light curves with Doppler boost variability plus a DRW component: 
\begin{equation}
\label{eq:V_M2_models}
  V=V_{\rm DB}+V_{\rm DRW} \quad\text{and}\quad M2=M2_{\rm DB}+M2_{\rm DRW}
\end{equation}
We first assume the simplest model for relativistic Doppler boost, i.e. constant luminosity in the rest frame of the SMBH, which gives rise to a sinusoidal light curve.
Therefore, $V_{\rm DB}=A_V \sin(2 \pi t/P+\phi)$, and $M2_{\rm DB}=A_{\rm DB}\times V_{\rm DB}$, where $A_{\rm DB}=(3-\alpha_{M2})/(3-\alpha_V)$. For our fiducial model, we set  $P=1994$\,d, $A_V=0.14$\,mag, $\phi=\pi$ and $A_{\rm DB}$=2.17, following \citetalias{Dorazio2015Nature}.

For the DRW light curves, we use the power spectral distribution from \citet{Kozowski2010},
\begin{equation}
PSD(f) = \frac{2\sigma^2{\tau}^2}{1+(2\pi f \tau)^2} \label{eq:3},
\end{equation}
with $\sigma=0.071\,\text{mag}/\sqrt{d}$ and $\tau=48$\,d from \citetalias{Dorazio2015Nature}.\footnote{As mentioned above, 
the best-fit DRW parameters of PG1302-102 reported in the literature have a wide range. 
 Below we explore a range of values that cover the published results.}
Using the prescription from \citet{Timmer1995}, numerically implemented in python in the astroML package \citep{astroML,astroMLText}, we generate evenly sampled DRW time series with a cadence of 1\,d. We downsample the data at the observed times and add Gaussian errors, with zero mean and standard deviations equal to the photometric errors, in order to generate light curves with properties similar to the observations.
We assume that the DRW model has similar amplitudes in optical and UV, i.e. $\sigma_{\rm opt}=\sigma_{\rm UV}$ (but relax this assumption below). 
We generate a distribution of $R_{\rm ij}$ by simulating 1,000 mock light curves. 

We test the null hypothesis (i.e. the multi-wavelength light curves are consistent with the relativistic Doppler boost plus DRW model) by examining whether the distribution of $R_{\rm ij}$ from the observed light curve $R_{\rm obs}$ is drawn from the same distribution as the simulated data $R_s$. 
Typically, a Kolmogorov-Smirnov (KS) test is performed.
However, the KS test assumes that the measurements are independent and identically distributed, which is not true for $R_{\rm ij}$, since we use multiple pairwise combinations of the same data points.

We overcome this limitation by employing the basic principle of the KS test, while accounting for the fact that the values $R_{\rm ij}$ are not independent. More specifically, the KS test quantifies the difference between a sample and a reference distribution with the maximum distance $\mathcal{D}$ between the empirical distribution function (EDF) of the sample and the cumulative distribution function (CDF) of the reference distribution. Confidence limits are then commonly obtained from approximate or asymptotic distributions of the distance $\mathcal{D}$ between independent realisations of the reference distribution. We here consider the distribution of $R_s$ from all simulated realisations as the reference distribution, and similarly to the KS test, we define $\mathcal{D}_{\rm obs}$, the maximum distance between the EDF of $R_{\rm obs}$ and the CDF of $R_s$ as our test statistic. However, we then explicitly compute the null distribution (i.e. the distribution of the test statistic) from the simulated data by calculating the maximum distance $\mathcal{D}_s$ between the EDF of each realisation and the CDF of $R_s$.

\begin{figure}
 \includegraphics[width=\columnwidth]{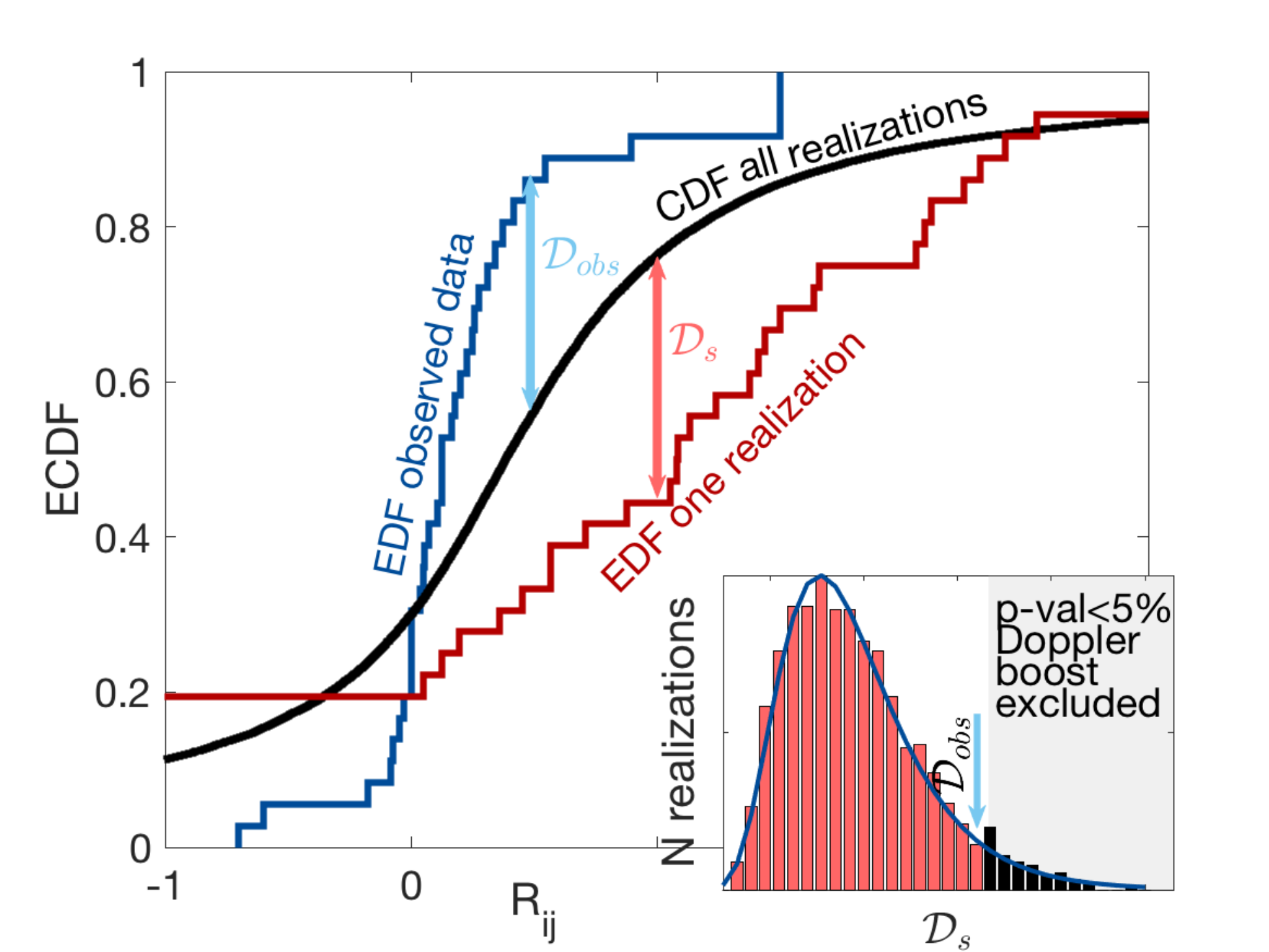}
 \caption{Illustration of the test statistic and the calculation of the $p$-value of the null hypothesis.}
 \label{fig:method}
\end{figure}

We define the $p$-value as the fraction of realisations that have maximum distance $\mathcal{D}$ greater than the observed ($\mathcal{D}_s>\mathcal{D}_{\rm obs}$). Note that a small $\mathcal{D}$ value indicates good agreement between the sample and the reference distribution. If the $p$-value is less than 5\%, we can reject the null hypothesis at the 5\% level. If, on the other hand, the $p$-value is greater than 5\%, the evidence against the null hypothesis is weak, and the observations could be consistent with relativistic Doppler boost. In Figure~\ref{fig:method}, we illustrate the test statistic and the calculation of the $p$-value. In the completely idealised case (without any extra intrinsic variability (e.g., DRW) from the quasar and without photometric errors), the CDF would be a step function at 0.46 (1/2.17).

Finally, we explore how the choice of parameters (namely, the DRW parameters $\sigma$ and $\tau$, the Doppler boost amplitude ratio $A_{\rm DB}$ 
and the relative amplitude of the intrinsic quasar variability in optical and UV $\sigma_{\rm opt}/\sigma_{\rm UV}$) affect our results. Specifically, we vary $\sigma$ from 0.003 to 0.08 mag/$\sqrt{d}$ and $\tau$ from 30 to 500 $d$ on a 10$\times$10 linearly spaced grid to sufficiently cover the best-fitting parameters in Table~\ref{Table:DRW_BestFit}, due to the their uncertainties.

\section{Results}
\label{sec:results}

\subsection{Doppler boost test in the V and M2 bands}
We test the multi-wavelength Doppler boost signature for the binary candidate PG1302-102 with simultaneous optical and UV observations. 
For direct comparison with \citetalias{Dorazio2015Nature}, we first test the light curves in $V$-band (optical) and $M2$-band (UV)\footnote{In \S~\ref{sec:morebands}, we extend the test to additional bands.} adopting the parameters from their analysis.
For the fiducial model, we find that only 47 realisations produce a maximum distance larger than the observed ($p$-value=4.7\%) and thus we can reject the  hypothesis that the data are consistent with the Doppler boost model. 

As stated in \S~\ref{sec:intro}, there is significant uncertainty with respect to the best-fitting DRW parameters for PG1302-102. 
For reference, we also calculate the expected values of $\sigma$ and $\tau$ for a typical quasar with the luminosity and redshift of PG1302-102. For this, we use the equations from \citet{2010ApJ...721.1014M}
\begin{equation} 
\log(X) = A+B\log\bigg(\frac{\lambda/(1+z)} {4000\text{\AA}}\bigg) +C(M_i+23) 
 +D\log\bigg( \frac{M_{\rm BH}}{10^9 M_\odot}\bigg)
\label{eq:6}
\end{equation}
with (A, B, C, D)=(2.4, 0.17, 0.03, 0.21) for $X=\tau_{RF}$\\
and  (A, B, C, D)=(-0.51, -0.479,  0.131,  0.18) for $X=\sqrt{2}\sigma$.\\
For PG1302-102, $M_{\rm BH}=10^{9} {\rm M_{\odot}}$, $z=0.27$ and for $V$-band $\lambda=5402$\,\AA. The absolute $i$-band magnitude can be calculated directly from the optical/IR spectrum in \citet{Graham2015a}; we calculate the (rest-frame) $i$-band flux, $F_i = 2 \times 10^{-13} \text{ erg} \text{ cm}^{-2} \text{ s}^{-1} $ and convert it to an $i$-band luminosity and subsequently to absolute magnitude ($M_i = -23.2$). In Table~\ref{Table:DRW_BestFit}, we summarise the values of $\sigma$ and $\tau$ from previous studies, along with the estimated values from \citet{2010ApJ...721.1014M}.

The large range of reported best-fit DRW values shown in Table~\ref{Table:DRW_BestFit} are likely responsible for the controversy regarding the significance of PG1302-102's periodicity. The values are not directly comparable, because each study used different components of the light curve of PG1302-102 (e.g., \citealt{Vaughan2016} used only the CRTS data, \citealt{Liu2018} used CRTS+LINEAR, whereas \citealt{Charisi2015} and \citetalias{Dorazio2015Nature} analysed the full published light curve from \citealt{Graham2015a}) and somewhat different methods to constrain the DRW parameters.
For instance, \citet{Liu2018} binned the light curve in very wide bins, which can significantly affect both the parameter estimation and the periodicity significance, as demonstrated with simulated data by \citet{Zhu_2020}.
On the other hand, \citet{Vaughan2016} introduced a parameter to account for a bias in the photometric errors; using the light curve from \citet{Graham2015a} and without this extra parameter, the estimated DRW parameters were similar to \citet{Charisi2015} (S.~Vaughan; private communication).
Additionally,  as \citet{Kozlowski2017} demonstrated with simulated DRW light curves, it is particularly challenging to constrain the DRW parameters, especially when $\tau$ is relatively long compared to the baseline. 
Finally, the errors on the best-fit values (especially on $\tau$) are typically large. Therefore, the differences between the quoted best-fit values of the DRW parameters is not surprising.

\begin{figure}
 \includegraphics[width=\columnwidth]{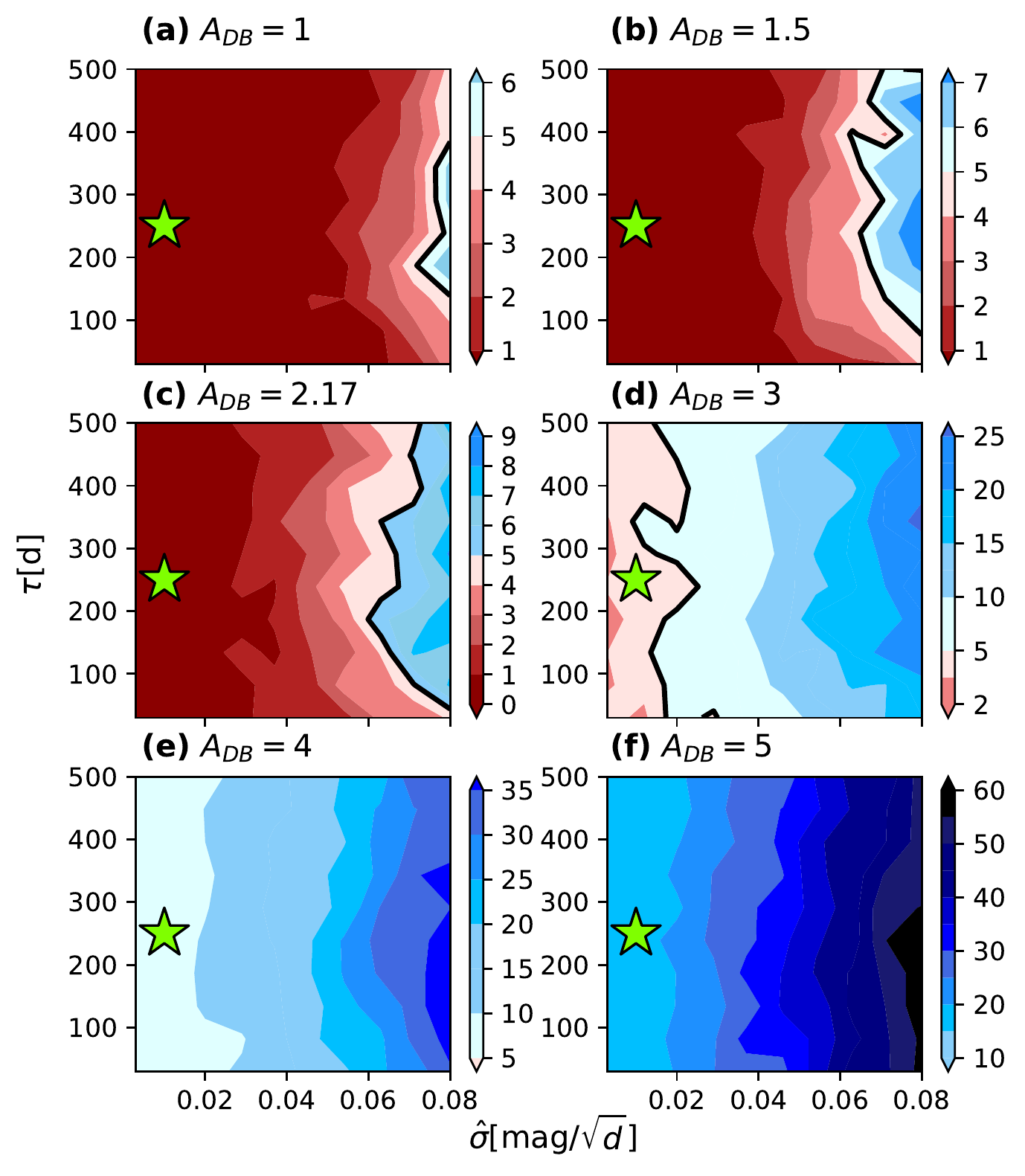}
 \caption{$P$-value (color bars; in unit of \%) of the null hypothesis (i.e. the multi-band light curves are consistent with Doppler boost plus DRW model) as a function of the DRW parameters $\sigma$ and $\tau$, with $A_{\rm DB}$ = 1 (top left), 1.5 (top right), 2.17 (middle left), 3 (middle right), 4 (bottom left) and 5 (bottom right) and UV/optical noise ratio $\sigma_{\rm UV}/\sigma_{\rm opt}=1$, considering the {\it Swift} $V$- and $M2$-bands. The black solid line represents the 5\% $p$-value threshold, which separates models that are rejected (red) from those passing the test (blue). The star corresponds to the average $\sigma$ and $\tau$ for quasars with properties similar to PG1302-102 from \citet{2010ApJ...721.1014M}.}
 \label{fig:V_M2_Doppler_test}
\end{figure}

Because of this, we take an agnostic approach and test the Doppler boost model for a wide range of DRW parameters. Initially, we keep the parameters of the Doppler boost model as in \citetalias{Dorazio2015Nature}.
In Figure~\ref{fig:V_M2_Doppler_test} (top left panel), we show the $p$-value of the null hypothesis as a function of $\sigma$ and $\tau$ for $A_{\rm DB}=2.17$. We see that for typical DRW parameters, we can reject that the data are consistent with the Doppler boost model, whereas for higher $\sigma$, the evidence against the Doppler model becomes weaker. 
In particular, at the fiducial $A_{\rm DB}=2.17$, the only allowed models are those with $\sigma\gsim 0.07~{\rm mag/\sqrt{d}}$. 
At face value, this indicates that the DRW component of variability dominates, and is a better description of the data. For example, the study by \citet{Vaughan2016}, which found the highest best-fitting values for the DRW parameters, questioned the significance of the periodicity; they concluded that the DRW model is preferred to a purely sinusoidal model. On the other hand, the DRW+sinusoidal model was found to be a better fit to the data than a pure DRW model in other studies~\citep{Dorazio2015Nature,Charisi2015,Liu2018}.
In our model, it is likely that a large $\sigma$ value is preferred, due to the lack of variability in the $V$-band. For instance, the amplitude of the $V$-band variability inferred solely from the {\it Swift}  observations is underestimated compared to the respective amplitude inferred from the ASAS-SN data, which can plausibly be attributed to unfortunate sampling (see \S~\ref{sec:OpticalAmplitude} for a detailed discussion).

\begin{table}
    \centering
    \begin{tabular}{l|l|l}
    Reference & $\sigma$ [mag/$\sqrt{d}$] &  $\tau$ [d] \\ \hline
    \citet{2010ApJ...721.1014M}     & 0.01 (0.15 mag)   & 245 \\
    \citetalias{Dorazio2015Nature}     & 0.071 (0.049 mag)  & 48\\
    \citet{Charisi2015} & 0.0157 & $\sim$100\\
    \citet{Vaughan2016} & 0.004 &550\\
    \citet{Liu2018}\footnote{These parameters were calculated for the light curve that includes only LINEAR+CRTS data points. With the inclusion of the ASAS-SN data, the best-fitting parameters are $\sigma$= 0.004\,mag/$\sqrt{d}$ and $\tau$=610d (T.~Liu; private communication)}& 0.005&429\\
    \hline
    \end{tabular}
    \caption{Best-fitting DRW parameters from previous studies.}
    \label{Table:DRW_BestFit}
\end{table}

Subsequently, we investigate how the assumptions in the Doppler boost model affect our results. First, we vary the amplitude ratio and repeat our tests for $A_{\rm DB}=1,1.5,3,4,5$.\footnote{Technically, the relative amplitude is not an assumption, but a robust prediction of the Doppler boost model if the spectral indices are known. However, it is possible that the spectral slopes vary, leading to varying values of the Doppler boost amplitude. Furthermore, if the amplitude in one band is poorly constrained, e.g., see \S~\ref{sec:OpticalAmplitude}, the estimate of the relative amplitude can be significantly affected.} In (part of) Figure~\ref{fig:V_M2_Doppler_test}, we show the $p$-value as a function of $\sigma$ and $\tau$, for $A_{\rm DB}=3$ (top right), $A_{\rm DB}=4$ (bottom left), and $A_{\rm DB}=5$ (bottom right). For higher Doppler boost amplitudes, we cannot reject the Doppler boost hypothesis, except for a small range of $\sigma$ and $\tau$, in the case of $A_{\rm DB}=3$. Nevertheless, the UV/optical spectral slopes of PG1302-102, estimated in \S~\ref{sec:morebands}, are in tension with such a high Doppler boost amplitude ratio. Additionally, we also explore smaller $A_{DB}$ values (1 and 1.5) in Figure~\ref{fig:V_M2_Doppler_test} because from the control sample in \citet{Charisi2018}, we see that the expected values of this ratio range from $A_{\rm UV}/A_{\rm opt}=1-2$ for typical spectral slopes.  However, in \S~\ref{sec:OpticalAmplitude} we discuss a potential explanation for the high value required to pass our statistical test.

\begin{figure}
 \includegraphics[width=\columnwidth]{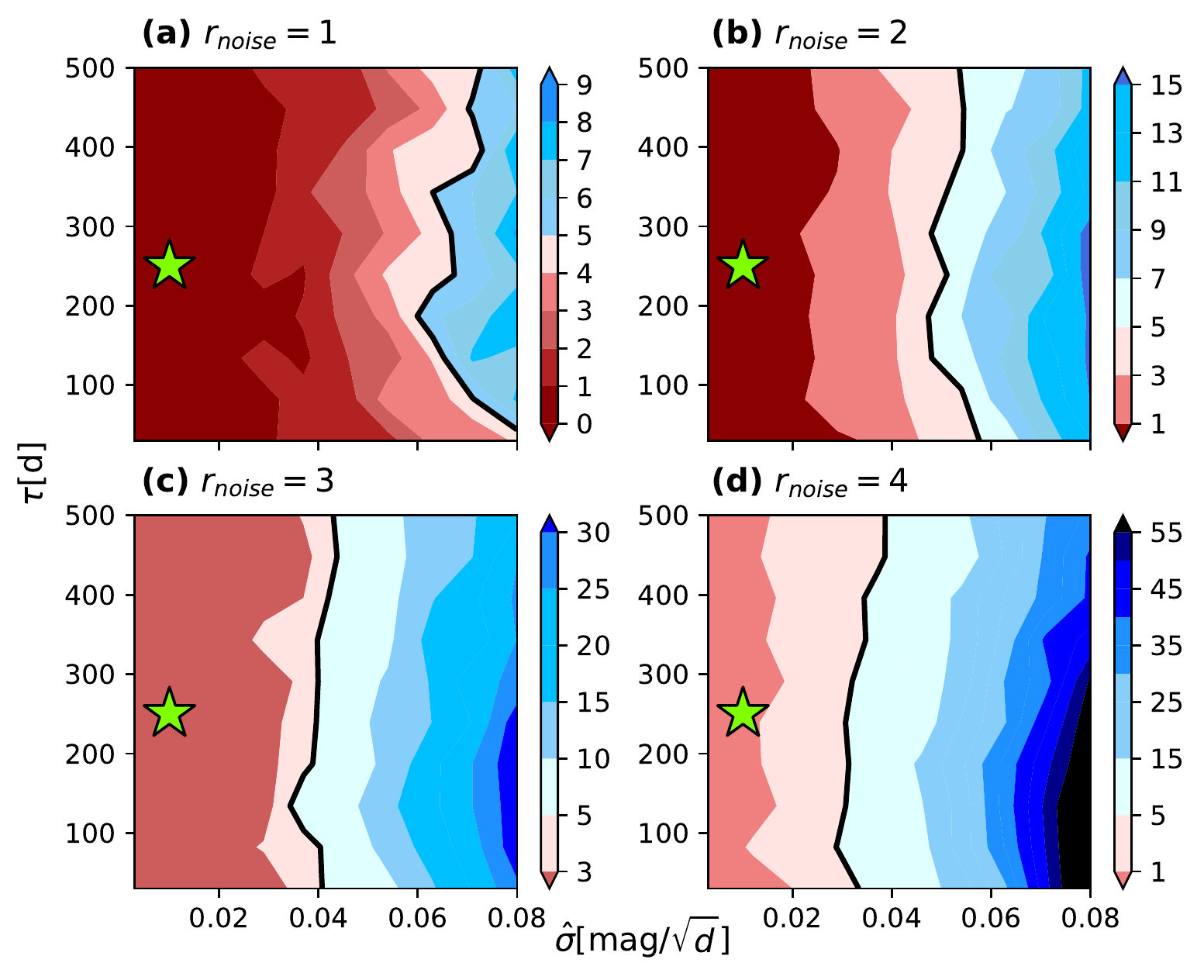}
 \caption{$P$-value (color bars; in unit of \%) as a function $\sigma$ and $\tau$, for $A_{\rm DB}=2.17$ and
 with $r_{\rm noise}=\sigma_{\rm UV}/\sigma_{\rm opt}=1$ (top left), 2 (top right), 3 (bottom left) and 4 (bottom right), again considering the {\it Swift} $V$- and $M2$-bands.} 
 \label{fig:uv_opt_noise_params}
\end{figure}

An additional assumption in our fiducial model is that the DRW variability in the the optical and UV bands have the same amplitude ($\sigma_{\rm UV}=\sigma_{\rm opt}$; note that in the simulations, the DRW light curves are drawn independently in the two bands, although 
see also \S~\ref{sec:multiband} and \ref{sec:limitations}). However, there is significant evidence that quasars have wavelength-dependent variability, with higher amplitudes at shorter wavelengths (e.g., \citealt{VandenBerk2004,Welsh2011}), with the variability in optical and UV bands correlated (\citealt{Hung2016,Buisson2017}).
Motivated by this, we increase the amplitude of the DRW variability in the UV to reflect the intrinsic colour-variability of quasars. We repeated our tests of the Doppler boost hypothesis for $r_{\rm noise}\equiv \sigma_{\rm UV}/\sigma_{\rm opt}=2,3,4$; the $p$-value in each case is shown in Figure~\ref{fig:uv_opt_noise_params}. This figure shows that as the relative amplitude of the DRW is increased, the Doppler boost model is excluded for a smaller range of parameters. However, this effect is relatively less significant.

\subsection{Test in other bands}
\label{sec:morebands}
Since our observations with {\it Swift}/UVOT cover six distinct bands, we extend the test to additional combinations of bands (in particular, we test the model in $B$ versus $M2$ and $V$/$B$ versus $W1$). For this, we first calculate the spectral slopes in the remaining bands (beyond $M2$ and $V$), to predict the expected relative amplitudes $A_{\rm DB}$ from eq. (\ref{eq:1}).

\begin{figure}
 \includegraphics[width=\columnwidth]{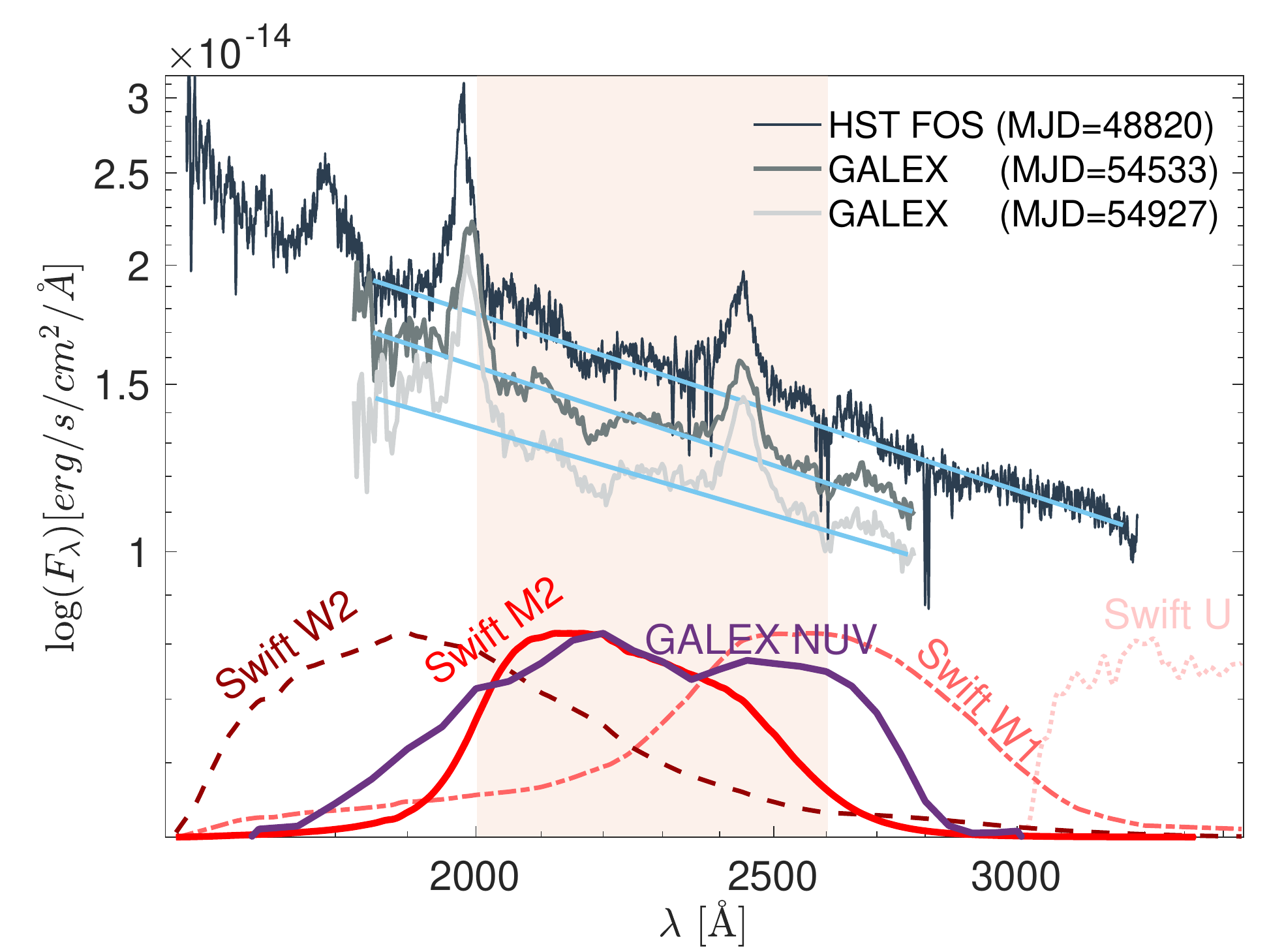}
 \caption{UV spectra from HST (black) and GALEX (grey and light grey). The blue lines show power-law fits to the continuum. The transmission curves of the {\it Swift}/UVOT (and GALEX NUV) filters are shown to delineate the wavelength coverage of each band --- the transmission curves are for illustrative purposes and are not measured in flux units ($y$-axis). The shaded band indicates the wavelength range in which the near-UV spectral slope was estimated in \citet{Dorazio2015Nature}.}
 \label{Fig:UV_Spec}
\end{figure}

In Figure~\ref{Fig:UV_Spec}, we show the UV spectra, presented in \citetalias{Dorazio2015Nature}, focusing on the wavelength range covered by the four UV bands of {\it Swift}. For reference, we also show the transmission curves of the {\it Swift} filters and the {\it GALEX} near-UV filter. There are three available UV spectra (one from {\it HST} and two from {\it GALEX}), taken several hundred days apart. We fit a power-law to the continuum $F_{\lambda}\sim \lambda^{\beta_{\lambda}}$ and  calculate the spectral index $\alpha_{\nu}=-\beta_{\lambda}-2$ (Table \ref{Table:spectral_slopes}). Our estimate of $\alpha_{\nu}\sim-1$ is in agreement with the value in \citetalias{Dorazio2015Nature}. In Table \ref{Table:spectral_slopes}, we show the exact values obtained from each spectral fit. Figure~\ref{Fig:UV_Spec} shows that a single power-law can reasonably describe the continuum of PG1302-102 in almost all UV bands. We also see that the UV spectral index does not change significantly over time.  Additionally, the available spectra cover only a small fraction of the $U$ band. Therefore, we exclude this band from the analysis, since we cannot estimate the spectral index. We also exclude $W2$, because it significantly overlaps with the broad CIV line; in SMBHBs, the broad emission lines are unlikely to be associated with the mini-discs \citep{Lu+2016}, but its presence in the wavelength range of $W2$ may lead to additional variability, which is not related to Doppler boosting.

\begin{figure}
 \includegraphics[width=\columnwidth]{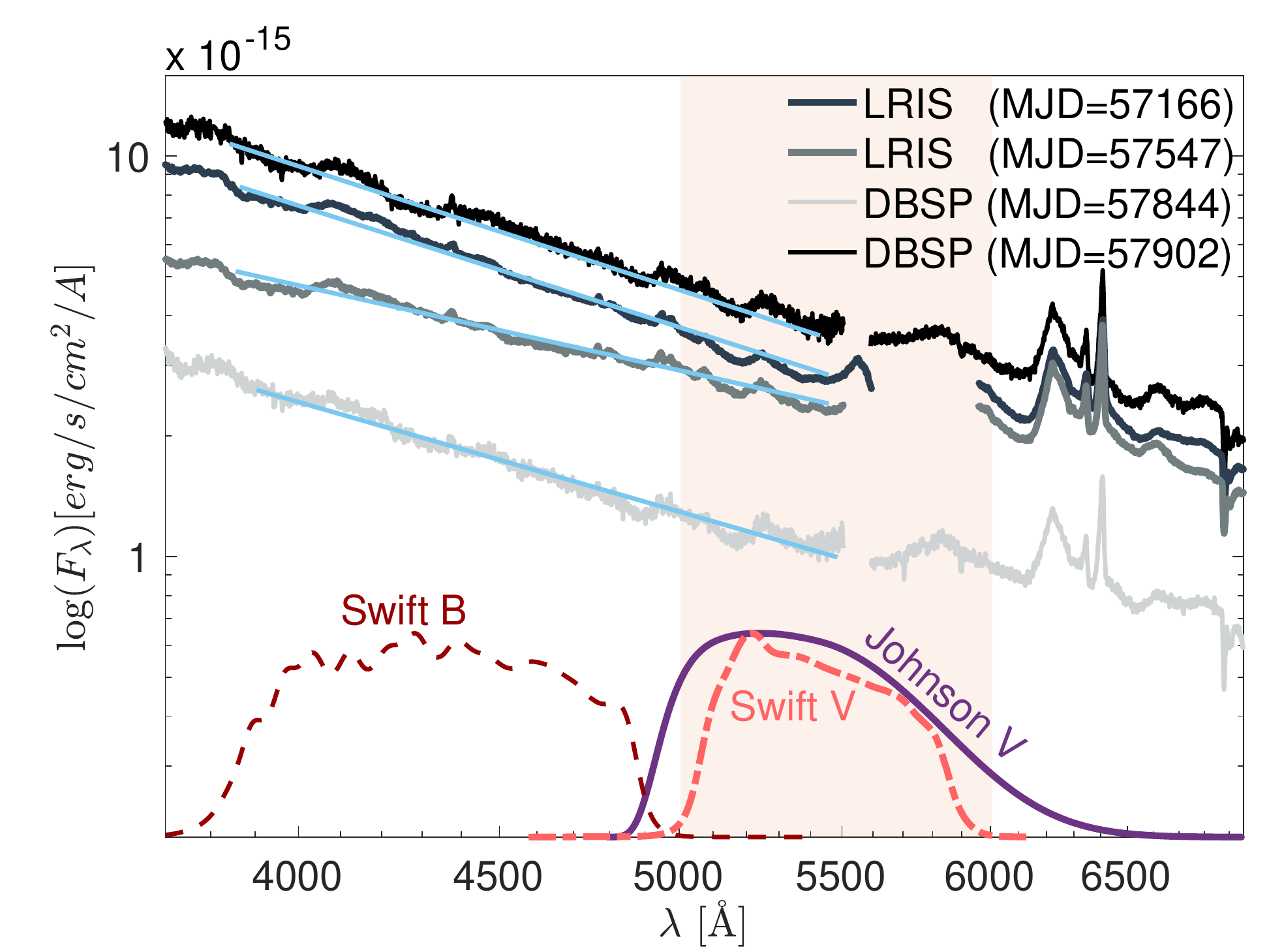}
 \caption{Optical spectra taken with DBSP at Palomar and LRIS at Keck.}
 \label{Fig:Opt_Spec}
\end{figure}

As part of an ongoing effort to spectroscopically follow up the SMBHB candidates from CRTS \citep{Graham2015b}, we have obtained four optical spectra of PG1302-102, two with the Low Resolution Imaging Spectrometer (LRIS) on the WM Keck Observatory and another two with the Double Spectrograph (DBSP) on the Palomar 200 inch telescope. 
In Figure ~\ref{Fig:Opt_Spec}, we show the optical spectra (along with the transmission curves of the optical filters of UVOT and Johnson $V$-band), with a power-law fit to the continuum. We summarise the estimated spectral slopes in Table \ref{Table:spectral_slopes}.
The continuum in optical bands (from $\sim$3800\,\AA\, to $\sim$5500\,\AA) can be successfully described by a single power-law.  For longer wavelengths (>5500\,\AA), the flux density $F_{\lambda}$ seems to flatten, consistent with the composite quasar spectrum from \citet{VandenBerk2001}. However, because of the gap between the blue and red channels of DBSP and LRIS, the calibration of the two spectral components is slightly uncertain and thus the estimation of the slope in this part of the spectrum is challenging. We consider the value obtained from fitting the blue component of the spectrum to be an upper limit for the spectral slope in the $V$-band. With the exception of the spectrum taken on MJD=57547, the spectral index is roughly constant. However, since the spectral slope in one of the four spectra is significantly different, we cannot exclude the possibility that the spectral index may vary over time.

\begin{table}
    \centering
    \begin{tabular}{l|l|l|l}
    & MJD & $\beta_{\lambda}$ & $\alpha_{\nu}$ \\ \hline
    \multirow{4}{*}{UV} 
    & 48820   & -1.05 & -0.95 \\
    & 54533 & -1.07 & -0.93\\
    & 54927 & -0.95 & -1.05\\
    & \citetalias{Dorazio2015Nature} & -0.95 & -1.05\\
    \hline
    \multirow{5}{*}{Optical}&57166&-3.13&1.13\\
    &57547&-2.19&0.19\\
    &57844&-2.83&0.83\\
    &57902&-3.19&1.19\\
    &\citetalias{Dorazio2015Nature}&-3.10&1.10\\
    \end{tabular}
    \caption{UV and optical spectral slopes from fitting the continuum with a single power-law.}
    \label{Table:spectral_slopes}
\end{table}

\begin{figure}
 \includegraphics[width=\columnwidth]{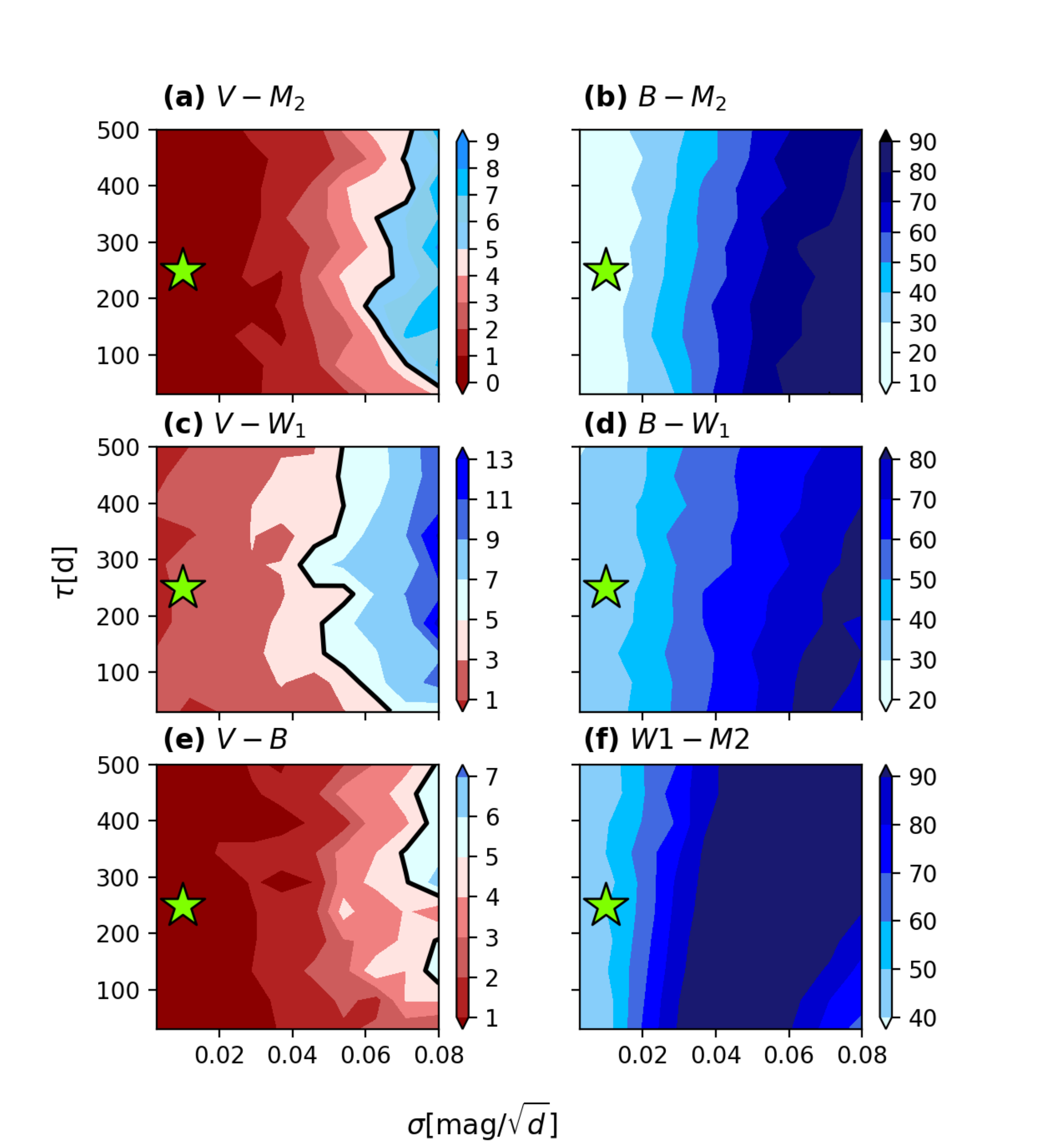}
 \caption{$P$-value (color bars; in unit of \%) as a function of DRW parameters $\sigma$ and $\tau$ for six combinations of independent UV ($M2$ and $W1$) and optical ($V$ and $B$) bands, where $A_{\rm DB}$ = 2.17 in panels (a)--(d) and $A_{\rm DB}$ = 1 in (e) and (f), and UV/optical noise ratio $r_{\rm noise}=1$.}
 \label{fig:Combination-of-bands}
\end{figure}

The spectral fits in Figure~\ref{Fig:Opt_Spec} show that the same power-law can describe the continuum both in $V$ and $B$ bands (although the continuum appears to flatten somewhat on the long-wavelength side of the $V$-band). From Figure~\ref{Fig:UV_Spec} we find that the spectral index is similar for $M2$, $W1$ and $W2$. Therefore, we can extend the test of the Doppler boost hypothesis to other combinations of optical/UV bands, with Doppler boost amplitude ratio $A_{\rm DB}=2.17$ for combinations of UV--optical bands ($V$--$M2$,$B$--$M2$,$V$--$W1$,$B$--$W1$) and $A_{\rm DB}=1$ for optical--optical ($V$--$B$) and UV--UV bands ($M2$--$W1$), where $r_{\rm noise}=1$ throughout this analysis for simplicity. In Figure~\ref{fig:Combination-of-bands}, we show the $p$-value for the following combinations of optical \& UV filters: $V$ versus $M2$ (top left), $B$ versus $M2$ (top right), $V$ versus $W1$ (middle left) and $B$ versus $W1$ (middle right). We see that,  when the $B$-band is considered, the data are consistent with the Doppler boost model for all the examined values of $\sigma$ and $\tau$.  In the bottom row, we show the results of the same Doppler ratio test, but applied internally within the optical $V$ versus $B$ (bottom left) and UV $W1$ versus $M2$ bands.  These show that the Doppler models are strongly ruled out because of the internal inconsistency within the optical bands with this model.
Note that because of the flattening of the $V$-band continuum spectra on the long--$\lambda$ side covering the $V$-band, the Doppler boost ratio is expected to be higher when $V$-band is involved, compared to the $B$-band.

\begin{figure*}
    \centering
    \includegraphics[width=\textwidth]{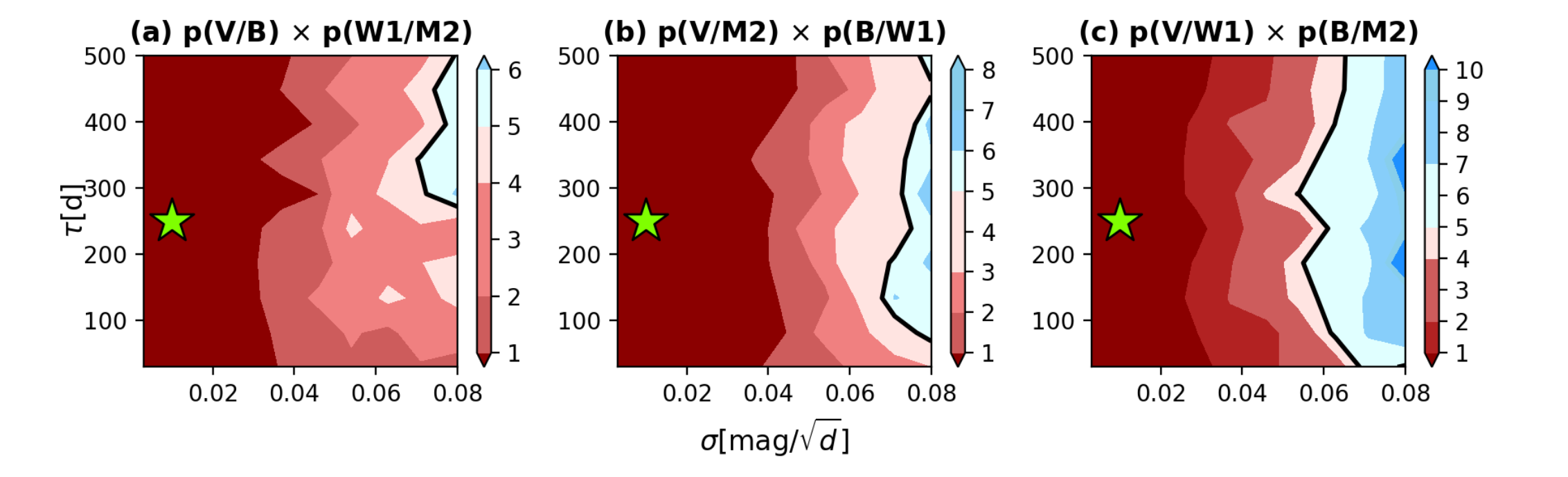}
    \caption{$P$-values (color bars; in unit of \%) of three pairs of independent combinations of bands {\it vs} the DRW parameters, assuming Doppler amplitude $A_{\rm DB}$ = 1 for the combinations of bands in panel (a) and $A_{\rm DB}$ = 2.17 in panels (b) and (C), and UV/optical noise ratio $r_{\rm noise}=1$.
    }
    \label{fig:Multi-band}
\end{figure*}

So far, we calculated the $p$-value of the Doppler boost hypothesis using the ratio test in all possible (six) pairwise combinations of the four bands, as shown in Figure~\ref{fig:Combination-of-bands}.
In order to assess the Doppler boost hypothesis in all bands simultaneously, we assume that the ratio tests are independent (if a specific band is not repeated). Therefore, there are three combinations of pairs, which cover all the bands without repetition; we can calculate the $p$-value of the full multi-band test by multiplying the $p$-values of the pairwise tests (e.g., p($V$/$B$)$\times$p($W1$/$W2$), etc.), as shown in Figure~\ref{fig:Multi-band}.
For example, the $p$-values in Figure~\ref{fig:Multi-band}(a) are the products of $p$-values in Figure~\ref{fig:Combination-of-bands}(e) \& (f). Figure~\ref{fig:Multi-band} shows statistical 
lower limits of the multi-band likelihoods of the Doppler model, because we assumed all bands are independent. 
These lower limits are between 1-2\%, implying that the model is ruled out at 98-99\% confidence.
In reality, the intrinsic variability of each band is correlated; accounting for these correlations should increase
the $p$-values compared to those shown in Figure~\ref{fig:Multi-band}. 

\section{Discussion}

\label{sec:discussion}

\subsection{Data extraction--pipeline caveat}
\label{sec:pipeline}
As mentioned in \S~\ref{sec:Data}, we extracted the {\it Swift} data using the on-line interface. Most epochs in the light curve (practically, all the epochs from our follow-up program in C13 and C14) consist of multiple observations taken very close in time. In this case, the on-line pipeline provides two options for magnitude estimation; the first relies on co-adding all the available images, whereas the second uses the image with the longest exposure time. 
We found that the data points extracted with the different options are not always in agreement. Since the individual exposures are not separated by long time intervals, it is unlikely that the observed discrepancy is caused by
quasar variability (e.g., see \citealt{Caplar2017}).

\begin{figure}
 \includegraphics[width=\columnwidth]{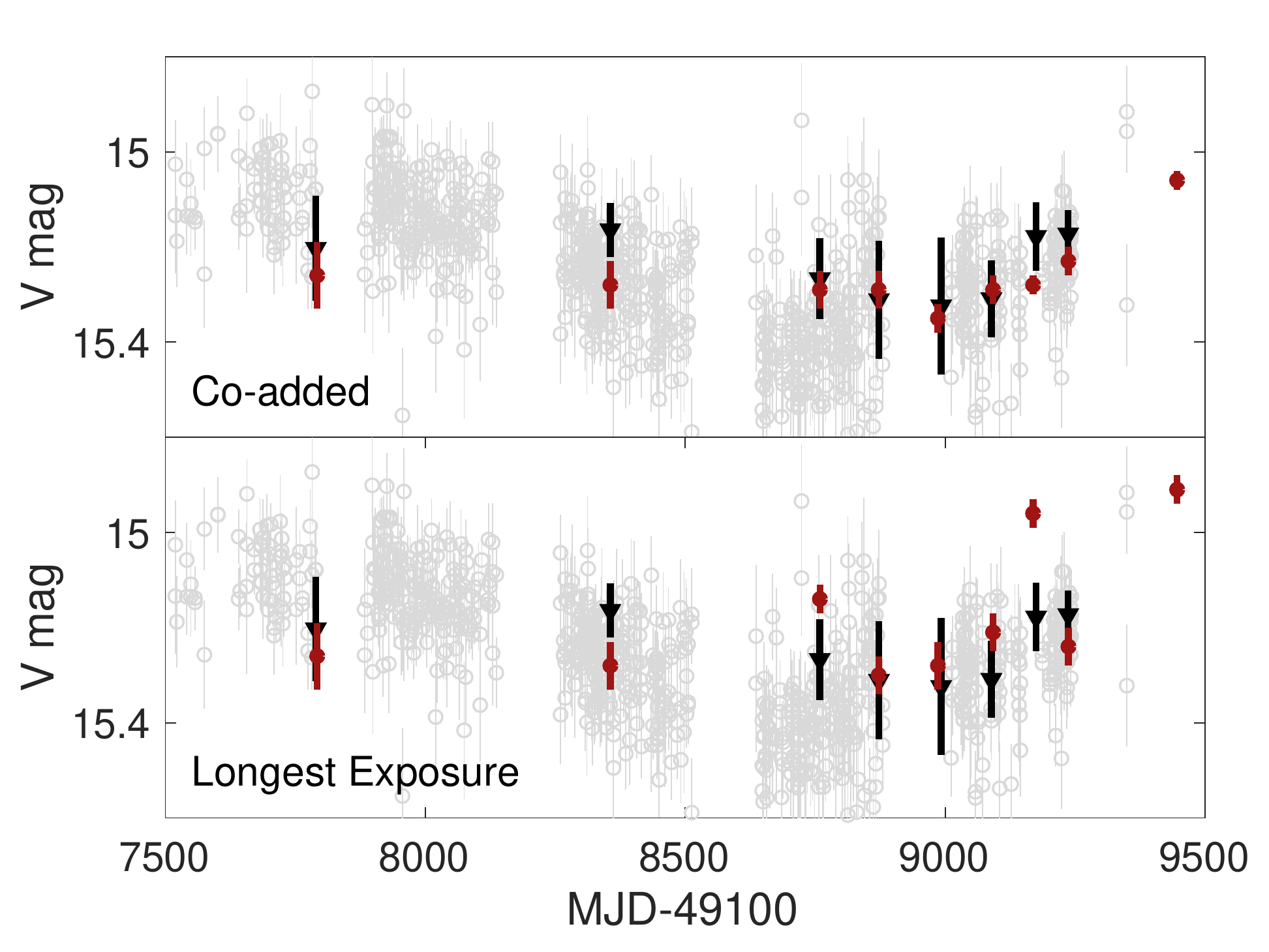}
 \caption{Comparison between ASAS-SN and {\it Swift} $V$-band photometry.  The ASAS-SN data points are identical in both panels and are shown in light grey, except for the points taken closest to the time of the nine {\it Swift} observations analysed in this paper, which are shown in black.  The red data points show {\it Swift} photometry from co-added images (top panel) and from the longest-exposure single image (bottom panel).  The co-added {\it Swift} data are in better agreement with ASAS-SN.}
 \label{Fig:Comparison_With_ASASSN}
\end{figure}

We devised the following tests to guide the selection of the optimal strategy for the data extraction. First, we cross-correlate the {\it Swift} observations with the light curve from ASAS-SN. As shown in Figure~\ref{Fig:Opt_Spec}, the $V$-band of {\it Swift} is very similar to Johnson $V$-band of ASAS-SN,\footnote{The photometry of ASAS-SN is calibrated with the APASS (AAVSO Photometric All-Sky Survey) catalogue, which was conducted also in Johnson $V$ (among seven other filters).} which allows for direct comparison. In Figure \ref{Fig:Comparison_With_ASASSN}, we show the ASAS-SN light curve superimposed with the {\it Swift} observations. We present the output light curves for the two data reduction options (co-added versus longest exposure, on the top and bottom panels, respectively),
see Table~\ref{Table:Swift_data}.  We highlight with black triangles the ASAS-SN observations that are closest in time to the {\it Swift} points. With the exception of the last epoch, there are nearly simultaneous observations of PG1302-102 from ASAS-SN (maximum one week apart). We see that the magnitudes from the co-added images are consistent with the ASAS-SN magnitudes within the photometric uncertainty, whereas the magnitudes from the images with the longest exposures are not in good agreement. Note that for the first two epochs, there is a single exposure and the magnitudes are identical in both light curves. 

Additionally, we examined the magnitudes of nearby stars in the images of PG1302-102. We found that with co-addition the stars had almost constant magnitudes, as expected, which was not true when we opted for the longest exposure images. From the above tests, we concluded that the light curves from the co-added images are more appropriate for our analysis. Even though the reason for the discrepancy is unclear, a potential explanation is that single exposures are more susceptible to outliers. Therefore, we caution future users of the on-line pipeline about this caveat.

\begin{table}
    \centering
    \begin{tabular}{l|l|l|l|l}
     & MJD & $t_{\rm expo}$ & $V$ mag & $\sigma_{m}$ \\ \hline
    \multirow{9}{*}{Co-added} 
    & 56892  & 62  &  15.26  &  0.07  \\
    &  57456  & 61  & 15.28    &  0.05      \\ 
    &  57859   & 136  &  15.29  &   0.04     \\  
    & 57972    & 127 &  15.29   & 0.04       \\  
     &  \textbf{58086}    &  \textbf{214}    &  \textbf{15.35}    &  \textbf{0.03}       \\
   &  \textbf{58192}    &  \textbf{170}    &  \textbf{15.29}    &  \textbf{0.093}     \\
   &  \textbf{58269}    &  \textbf{496}    &  \textbf{15.28}    &  \textbf{0.02}     \\ 
   &  \textbf{58337}    &  \textbf{566}    &  \textbf{15.23}    &  \textbf{0.03}       \\ 
   &  \textbf{58546}    &  \textbf{520}    &  \textbf{15.06}    &  \textbf{0.02}     \\ \hline
    \multirow{9}{*}{Long expo} 
    &  56892    & 62   &   15.26    &   0.07    \\
    &  57456    & 61   &   15.28    &   0.05    \\ 
    &  57859    & 136  &   15.14    &   0.03    \\ 
    &  57972    & 127  &   15.30    &	0.04    \\
    &  \textbf{58086}    & \textbf{112}  &  \textbf{15.28}    &  \textbf{0.05}    \\ 
    &  \textbf{58192}    & \textbf{92}   &  \textbf{15.21}    &  \textbf{0.04}    \\ 
    &  \textbf{58269}    & \textbf{137}  &  \textbf{14.96}    &  \textbf{0.03}    \\ 
    &  \textbf{58337}    & \textbf{135}  &  \textbf{15.24}    &  \textbf{0.04}    \\ 
    &  \textbf{58546}    & \textbf{136}  &  \textbf{14.91}    &  \textbf{0.03}    \\
    \end{tabular}
    \caption{{\it Swift} $V$-band archival data for co-added and longest exposure measurements (corresponding to the datapoints shown in the top and bottom panels of Figure~\ref{Fig:Comparison_With_ASASSN}, respectively). The bold-faced rows high-light the observations with multiple exposures, where these two differ. The table shows observation date in MJD, exposure time ($t_{\rm expo}$) in seconds, $V$-band magnitude ($V$ mag) and magnitude errors ($\sigma_{m}$).}
    \label{Table:Swift_data}
\end{table}

\subsection{Optical variability amplitude}
\label{sec:OpticalAmplitude}

When we examined the data in the $V$ and $M2$ (or $W1$) bands, the null hypothesis tests lead us to conclude that we can exclude the fiducial Doppler boost model. The model is feasible only when the ratio of the Doppler-boost amplitudes in the two bands is high, but current estimates of the UV and optical spectral slopes are in tension with this high required ratio. On the other hand, the data are consistent with the Doppler boost model when we consider the $B$-band observations (both with $M2$ and $W1$). 

A potential caveat that could explain the preference for high $A_{\rm DB}$, when the test involves the $V$-band can be seen from a careful examination of the light curves in Figure~\ref{fig:1}. The amplitude of the UV variability is similar to that in \citetalias{Dorazio2015Nature}, whereas the $V$-band variability, calculated solely based on the {\it Swift} data, is significantly smaller. The best-fitting amplitude for a sinusoid with the period and phase from \citetalias{Dorazio2015Nature} is 0.32\,mag for the $M2$-band and $\sim$0.1\,mag for the $V$-band --- and is reduced to only 0.028\,mag if the last observation is omitted. On the other hand, from the ASAS-SN light curve, which covers the same time interval but with many more observations, the inferred amplitude of the sinusoid is 0.13\,mag, similar to \citetalias{Dorazio2015Nature}. We note that, in~\S~\ref{sec:pipeline}, we demonstrated that the {\it Swift} data points are generally consistent with the closest (in time) data points from ASAS-SN (see also top panel of Figure~\ref{Fig:Comparison_With_ASASSN}).

Given that we base our conclusions on the small number of data points taken by {\it Swift}, it is likely that our results are affected by unfortunate sampling exaggerated by short-term variability of quasars (which cannot be easily accounted for in our analysis, since we assess the statistical significance of the model by simulating sinusoids). The model may be rejected, because the test is unable to reproduce the UV variations, relative to the optical, which are particularly small and thus the need for larger UV/optical Doppler boost ratios (Figure~\ref{fig:V_M2_Doppler_test}), and/or larger UV noise amplitudes (Figure~\ref{fig:uv_opt_noise_params}).
We note further that the $B$ band variability may be a cleaner test of the Doppler boost than the $V$ band.  This is because of the following. In a simple toy model of the binary nucleus of PG1302-102, the thermal emission comes from three distinct regions: (1) the circumprimary minidisk, (2) the circumsecondary minisdisk, and (3) the circumbinary disk.  While the emission in the $B$ and the UV bands can safely be attributed to the circumsecondary minisdisk (with small contributions from the circumprimary disk), the $V$ band luminosity may receive a significant contribution from the circumbinary disk, because of the tidal truncation of the circumsecondary disk (see Extended Data Figure~1 in \citealt{Dorazio2015Nature}).   Since the circumbinary disk emission does not share the Doppler boost of gas bound to the secondary BH, it can add, effectively, a contribution that spoils the Doppler test.

We conclude that additional $V$-band data would be important, and could change our conclusions, rendering previously rejected models acceptable, or increasing the confidence at which the Doppler-boosted models can be ruled out. The former possibility is already demonstrated above, since the inclusion of the last observation leads to a significantly increased amplitude for the sinusoid (from 0.028 to $\sim$0.1). 
This means that even a small number of additional observations may have a profound effect in resolving the conflicting conclusions between $V$- and $B$-band tests. It is therefore crucial to continue monitoring PG1302-102 in optical and UV bands.

\subsection{Period of PG1302-102}
\begin{figure}
 \includegraphics[width=\columnwidth]{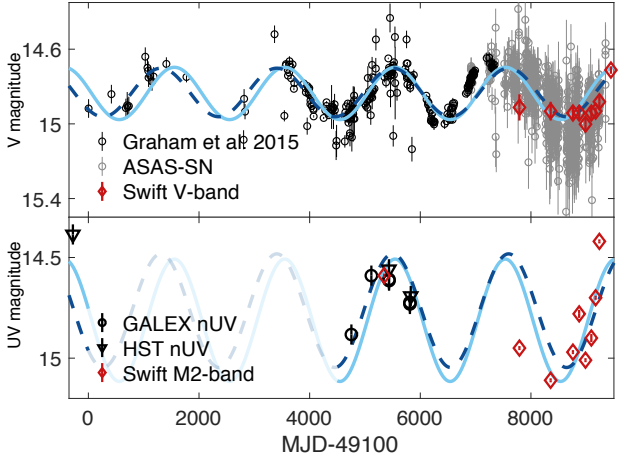}
 \caption{Optical $V$-band (top panel) and near-UV $M2$-band (bottom panel) light curve of PG1301-102, similar to Figure~\ref{fig:1}. The prediction of the Doppler model with parameters from \citetalias{Dorazio2015Nature} is shown with a solid light-blue line. The dashed dark-blue line represents a sinusoidal fit of the extended optical light curve, with period P=2095\,d longer than in \citetalias{Dorazio2015Nature}, and the corresponding Doppler boost prediction for the UV data with $A_{\rm DB}=2.17$ in the bottom panel.}
 \label{Fig:newPeriod}
\end{figure}

We tested the Doppler boost model for a fixed period and phase (P=1994\,d, $\phi=\pi$), using the parameters from \citetalias{Dorazio2015Nature}. This is consistent with \citet{Liu2018}'s estimate from a light curve including additional data from ASAS-SN ($P=2012.6^{+280}_{-220}$\,d).
We remind the reader that \citet{Graham2015a} had calculated a period of $P=1884\pm88$\,d, also consistent with both of the above estimates. 
Here we examine how the precise value of the period affects our results on the multi-wavelength Doppler boost signature. For this, we fit a sine wave to the extended light curve. We obtain a new best-fitting period of 2095\,d, which we show
in Figure~\ref{Fig:newPeriod} (dashed dark blue line), along with the best-fitting sinusoid from \citetalias{Dorazio2015Nature} (solid blue line), for comparison. We also present the rescaled sinusoids, which reflect the prediction of the Doppler boost model in UV, with $A_{\rm DB}=2.17$. We see that, with the updated period, the UV model does not fit the data equally well, because the UV points are slightly out of phase.

The estimated best-fitting period is within the one-sigma confidence intervals of \citet{Liu2018}. We note, however, that this sinusoidal fit is not directly comparable with \citet{Liu2018} for the following reasons:
1)~We fit a pure sine wave without a DRW component. 
2)~We consider the phase of the sinusoid as a free parameter and include it in the fit. 
3)~The light curve we analyse includes all the data points from \citet{Graham2015a} plus the ASAS-SN data, whereas \citet{Liu2018} examined only the data from CRTS, LINEAR and ASAS-SN.
4)~\citet{Liu2018} binned the entire light curve using wide bins of 150\,d ($\sim$100\,d for the ASAS-SN light curves and $180$\,d for the CRTS+LINEAR light curve),\footnote{This choice likely affects the DRW parameters rather than the sinusoid, because the bin size is much smaller than the period, but comparable to the expected $\tau$ parameter of the DRW.} but we analysed all the data points without binning. 
5)~We employ a non-linear regression, not an MCMC analysis.

\subsection{Deviations for sinusoidal variability}
Eq. (\ref{eq:deltaFnu}) predicts that, for a circular binary, the relativistic Doppler boost will produce smooth sinusoidal variations, if: A) the spectral indices are constant in both the optical and UV bands, and B) the rest-frame luminosity in the most luminous mini-disc is constant. If any of the above conditions is not met i.e., $\alpha_{\nu}\neq \text{const.}$ or $F_{\nu}\neq \text{const.}$, the variability will deviate from purely sinusoidal. Additional deviations may occur from the intrinsic quasar noise due to accretion onto the SMBHs, but this is taken into account with the addition of the DRW component in eq. (\ref{eq:V_M2_models}).

In our analysis, we assume that the ratio of the amplitudes $A_{\rm DB}$ is constant, i.e. the spectral indices remain unchanged through the long baseline of our observations. This seems to be a good assumption, especially in UV (Figure~\ref{Fig:UV_Spec}). In the optical bands (Figure~\ref{Fig:Opt_Spec}), however, there is some evidence for spectral variability, since the spectral slope changes significantly in one of the four optical spectra. Additionally, quasars have a well-established trend that their spectra become bluer in their brightest phase. Therefore, it is possible that the spectral indices vary over time. The statistical test we employed, based on the ratio of pairs of observations, cannot easily incorporate changes in $A_{\rm DB}$.

The null hypothesis test also assumes that the rest frame luminosity does not vary, since we assess the statistical significance of our findings by simulating sinusoids. Nevertheless, from hydrodynamical simulations, we expect fluctuations in the mini-discs, along with significant gas motion between the two SMBHs, which can contribute to additional Doppler boost variability beyond that from the orbital motion of the binary \citep{Tang+2018}.
In fact, the ratio test was designed in order to incorporate such changes. Unfortunately, these effects are not well understood, and it is thus particularly challenging to develop a physically motivated model to incorporate the additional variability in the statistical analysis. We recognize that this is an important effect, and we defer its addition to future work.

Furthermore, variations in the accretion rate may also produce optical and UV variability, which may be correlated, but not simultaneous.
While poorly understood even for single BHs, time-lags between the optical and UV variability are possible, and could strongly influence our analysis.
For example, if the accretion rate onto the binary is time variable, it may produce changes in flux that propagate inwards through the mini-discs (or the circumbinary disc), first causing a change in the optical band followed by a change in the UV. 
If the time-lag is of the order of the orbital or thermal time at $\sim 100 GM/c^2$, where typically the emission transitions from optical to UV, for PG1302-102 one would expect correlated variations on time-scales of months to years. If instead these flux variations are mediated by viscous processes, then the time-lag could be much longer, at least hundreds of years. Such a correlation between the UV and optical could help disentangle flux variations not induced by Doppler boost. 
Note that the expected time-lag in this binary scenario is the opposite of the time-lag in the regular accretion disc around a single SMBH, in which the optical follows the changes in the UV flux with minimal time-lags of days (at least from cross-correlations of two well-sampled light curves in \citealt{Buisson2017}).

\subsection{Wavelength-dependent Variability of Quasars}
\label{sec:multiband}
An important caveat in distinguishing the Doppler boost signature is that all quasars show wavelength-dependent variability, which can mimic the expected multi-wavelength Doppler variability, especially in the sparse data, which are typically available \citep{Charisi2018}.
This wavelength-dependent variability of quasars has been extensively characterized, and found to have larger amplitudes towards shorter wavelengths \citep{VandenBerk2004, MacLeod+2010, Schmidt+2012, Gezari+2013, Morganson+2014, Caplar2017}.  In particular, the recent study by 
\citet{Xin+2020} found that, in most quasars, the near-UV and optical variability are strongly correlated, with an amplitude ratio between $2\lsim A_{\rm UV}/A_{\rm opt}\lsim 3.5$.   Since this encompasses the value of 2.17 expected for the Doppler model given the optical/UV spectra of PG1302-102, a concern is that this ratio can arise by chance, from stochastic multi-wavelength variability.

In order to quantify how frequently the generic underlying colour-variability mimics the Doppler-induced value, it is necessary to compare the relative amplitude of optical and UV light curves 
$A_{\rm UV}/A_{\rm opt}$ with the expected value in the Doppler model, based on the spectral indices in the respective bands (see eq.~\ref{eq:1}), for a large control sample of aperiodic quasars.  This requires the availability of spectra in both bands, as well as optical and UV time-domain data (ideally sampled in both bands at the same time). Unfortunately, the number of quasars with such data is limited; \citet{Charisi2018} analyzed a small sample of 42 quasars and found that the Doppler signature can arise by chance in $\sim20\%$ (40\%) of the cases for the near-UV (far-UV) band. This probability reflects the limited data quality of the control sample (as is also demonstrated from the increase of chance coincidence in the lower-quality far-UV data), and represents only an upper limit on how frequently quasars mimic the Doppler color-variations. A larger sample of quasars with better UV+optical is needed to assess this caveat more accurately.
\vspace{\baselineskip}

\subsection{Constraints from Future Observations} \label{sec:forecast}

An interesting question to ask is how much the evidence for or against the Doppler model may tighten with continued multi-band monitoring of PG1302-102. To address this question, we generated hypothetical data representing future observations over $\sim 10$ years, covering two additional cycles of periodicity, and computed the $p$-values of the Doppler model with the extended data.  For simplicity, we focused on the $V$- and $M2$-bands only.

In particular, we assumed that the follow-up monitoring will continue with observations similar to those in Cycles 13 and 14. We mimicked the cadence of the existing {\it Swift} observations, by picking consecutive epochs of future observed dates that are 122 days apart (i.e. three epochs per year), and generated a total of 33 new mock observations to cover a baseline of 10\,yr. We also approximated the photometric errors for the mock $V$- and $M2$-band data by taking the average of the errors over the last seven {\it Swift} observations, which yields $\sim$0.03 and $\sim$0.01 mag, respectively.\footnote{For the estimation of the typical photometric error, we omitted the earliest two {\it Swift} observations, prior to our observing program in Cycles 13 and 14, because they have large errors due to short exposure times.} Finally, we combined the additional hypothetical data with the original 9 {\it Swift} points to construct the full new light curves.

\begin{figure*}
 \includegraphics[width=0.67\columnwidth]{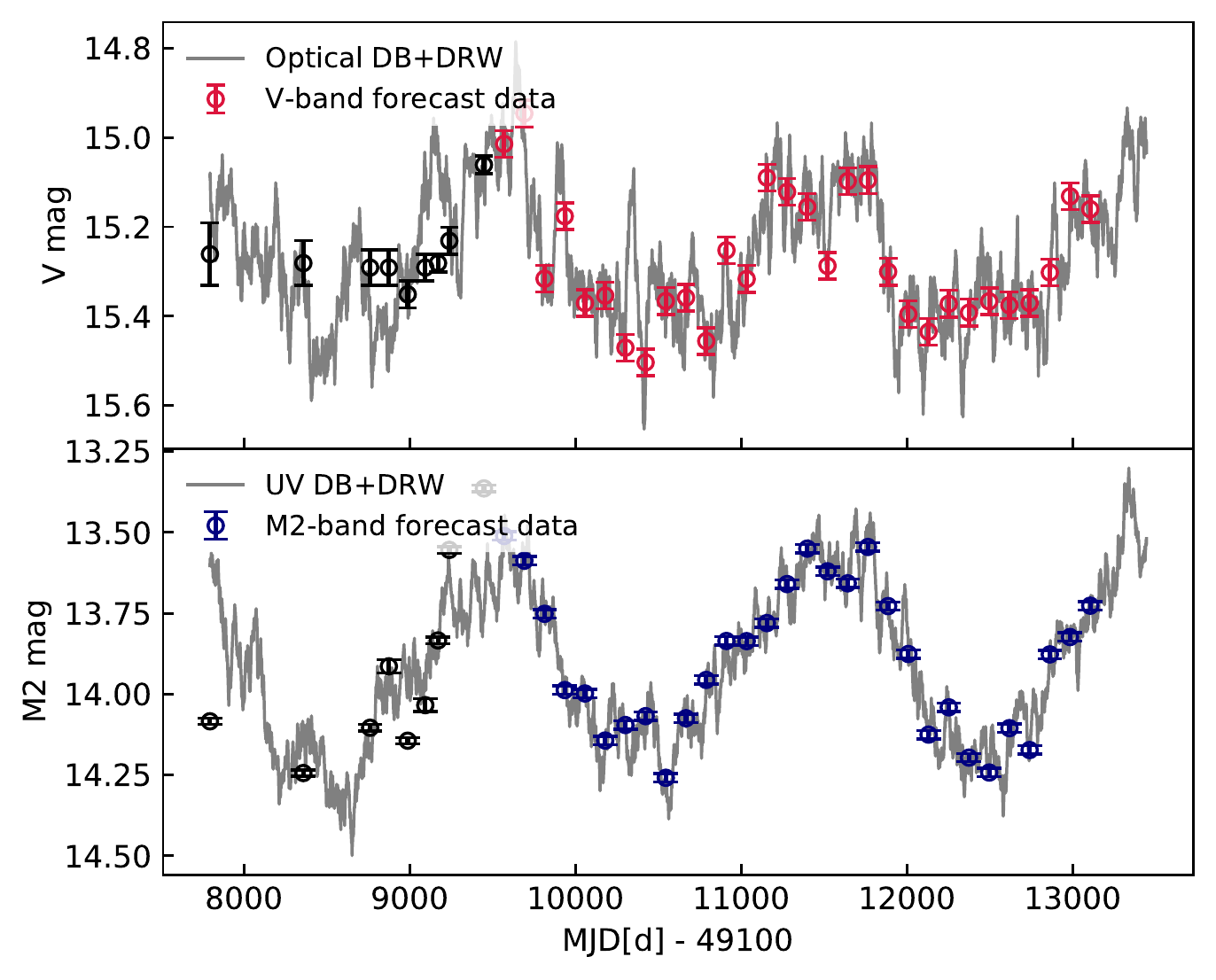}
  \includegraphics[width=0.67\columnwidth]{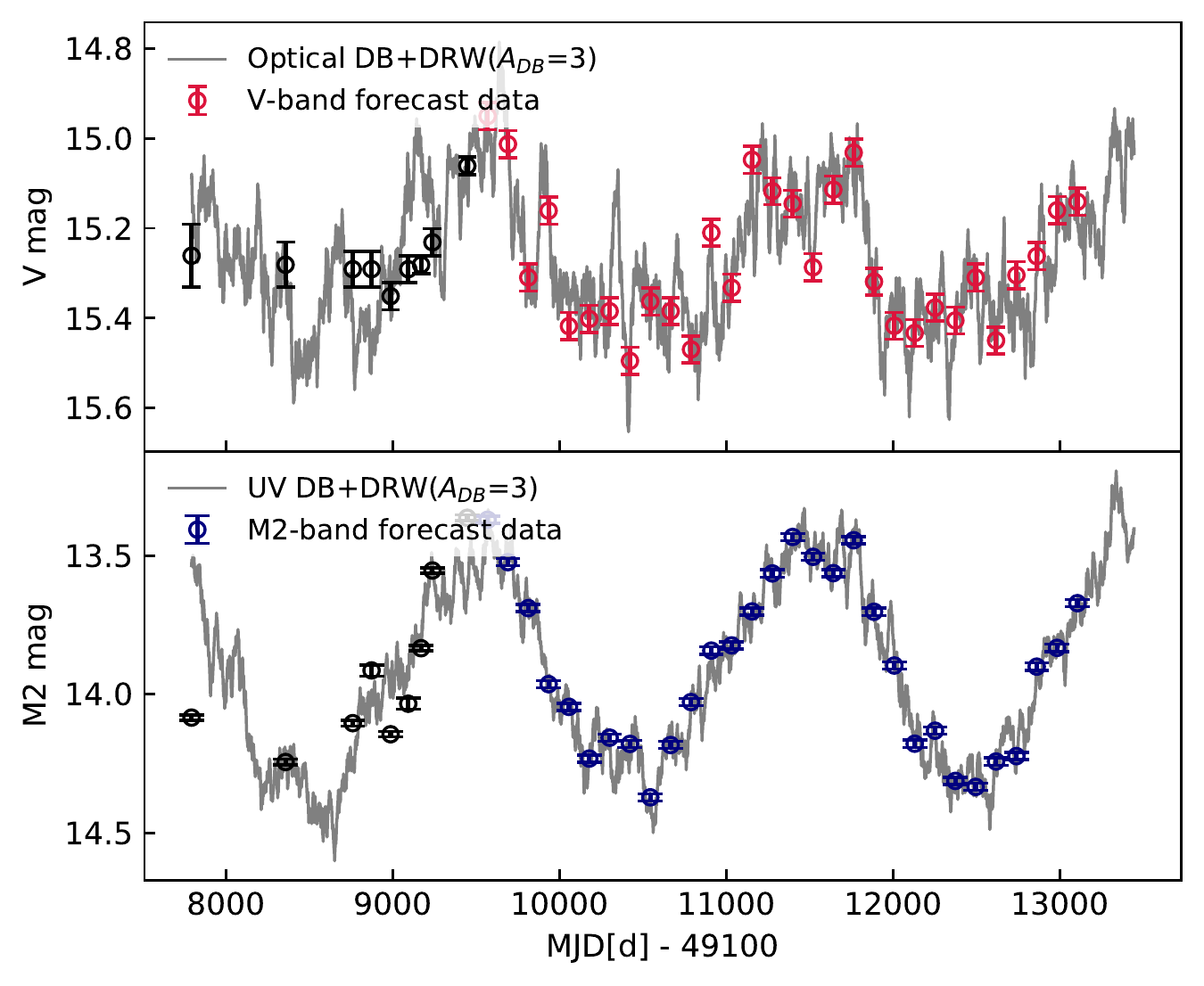}
  \includegraphics[width=0.67\columnwidth]{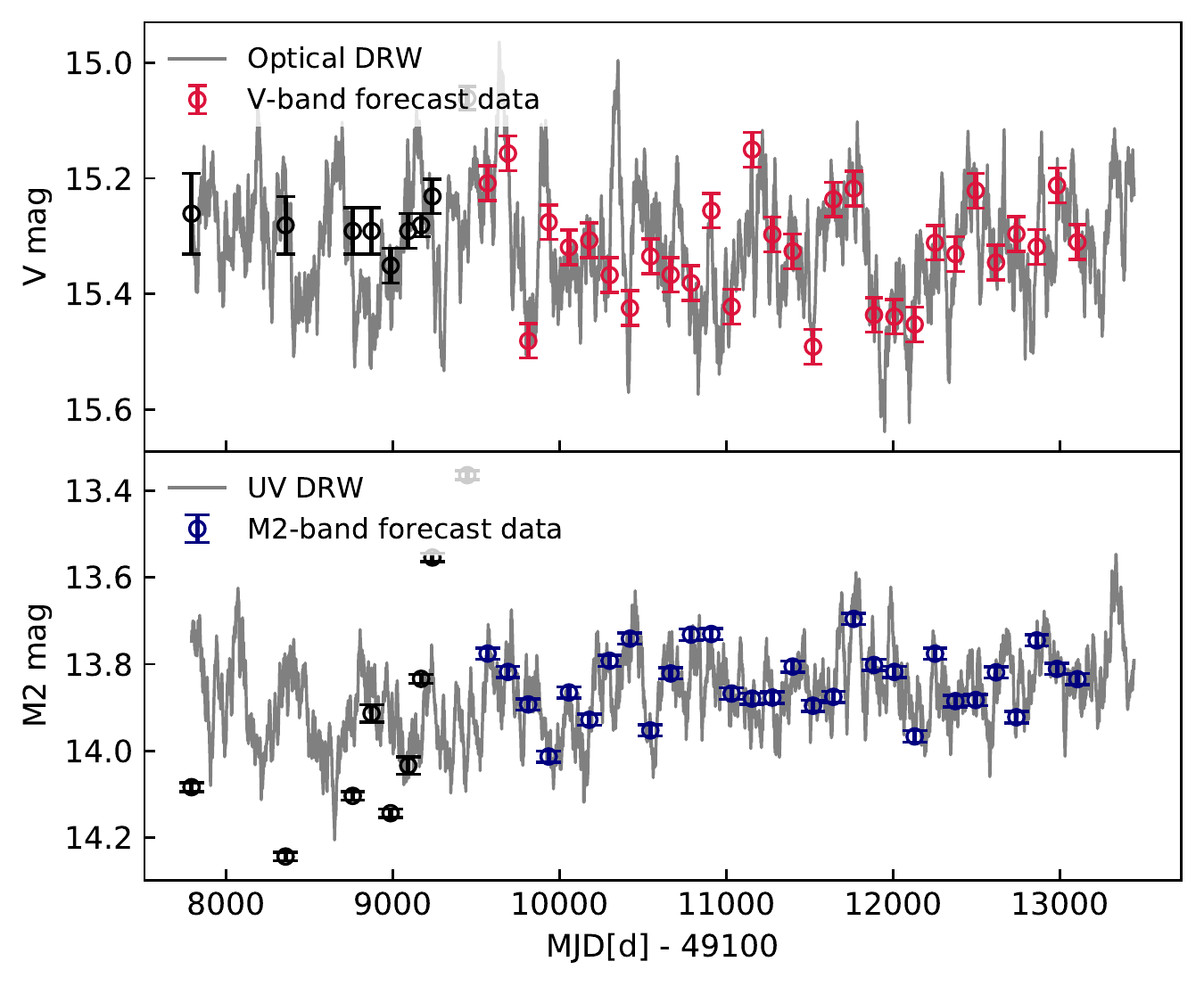}
 \caption{Example hypothetical light curves over upcoming $\sim$10 years, for each of the three cases  explained in \S~\ref{sec:forecast}; (1) the fiducial DB and DRW model with  $A_{\rm DB}$=2.17 dominates the variability of PG1302 (left); (2) DB + DRW model has higher Doppler amplitude ratio, $A_{\rm DB}$=3 (middle); and (3) DB is absent from the system and DRW model dominates the true variability (right). In all three scenarios, the DRW amplitude ratio is fixed at 1 (i.e. $r_{\rm noise}$=1). The black points with error bars are the existing 9 {\it Swift} observations in optical (top panels) and UV (bottom panels). The red and dark blue points with error bars are the hypothetical optical and UV data, respectively. } 
 \label{Fig:example_LCs}
\end{figure*}

\begin{figure}
 \includegraphics[width=\columnwidth]{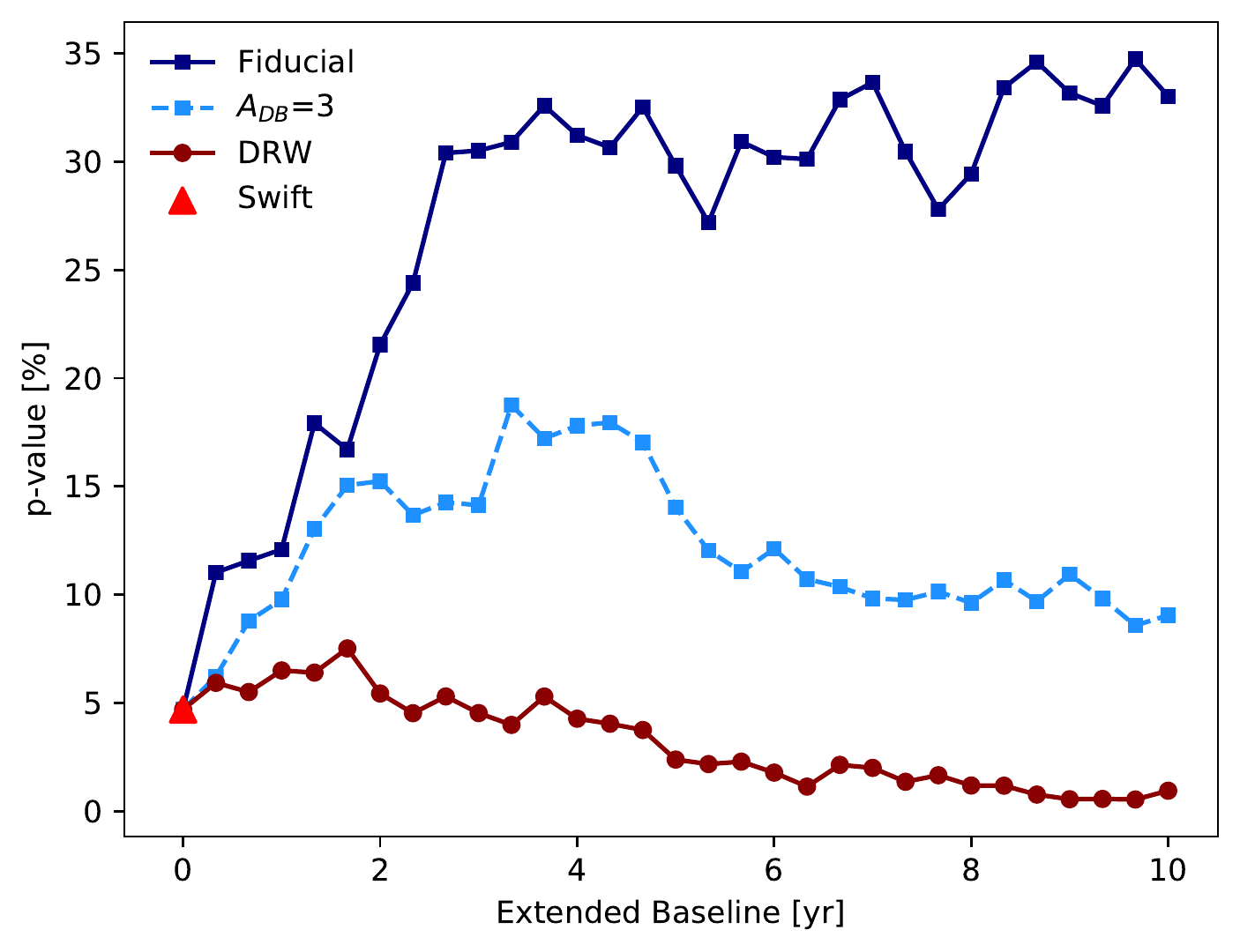}
 \caption{$P$-values inferred when the fiducial model is fit to three different hypothetical future datasets with extended baselines. The red triangle denotes the original 9 {\it Swift} points. The dark blue curve is inferred from mock data generated in the fiducial model itself. The light blue curve assumes that the true light curve has a larger Doppler amplitude ratio ($A_{DB}=3$). The dark red curve corresponds to future mock data consisting of pure DRW variability.}
 \label{Fig:forecast}
\end{figure}

First, we assumed that the Doppler boost model is, in fact, correct. We simulated 50 random realizations of mock $V$-and $M2$-band data, using the fiducial DB and DRW parameters from \citetalias{Dorazio2015Nature}.
We generated continuous DB and DRW light curves for a time period of 10\,yr,  down-sampled these at the forecast future MJDs and added photometric uncertainty. We show an example of a hypothetical light curve in the left most panel of Figure~\ref{Fig:example_LCs}. The other two panels in Figure~\ref{Fig:example_LCs} represent different models from the fiducial, which will be explained later in this section.
We then computed the corresponding $p$-values of the mock data, following the method described in \S~\ref{sec:Analysis}. 
We generated 1000 mock light curves to assemble the reference CDF, corresponding to the ``theoretical prediction" for the new data in the Doppler model (see Figure~\ref{fig:method}).  
In this analysis, we added the new mock data points one-by-one, and re-computed the $p$-values as each of the 1,2,...,33 new data points were added. For each added epoch, we calculated 50 $p$-values (one for each realization of the mock data). In Figure~\ref{Fig:forecast}, we show the average of the 50 $p$-values, as a function of the extension of the baseline (starting from the current constraints shown by the red triangle). The $p$-value for the DB+DRW model is shown by the top-most (dark blue) curve.

In the above exercise, we used the same model to generate and to fit the mock data - i.e. we assume that we correctly guessed the true nature of PG1302-102.  Under this assumption, the expected average $p$-value, as defined in our analysis, is $\langle p\rangle=50\%$;
we therefore expect the $p$-values to increase and approach this value as more mock data are added. The dark blue curve in Figure~\ref{Fig:forecast} shows this trend, although there are fluctuations during the additional 10 years, caused by the stochastic nature of the DRW and random photometric errors. Additionally, we note that despite a relatively steep rise over the first additional three years, the $p$-value reaches a plateau of $\sim 30\%$. These results suggest that data over an additional $\sim$1 cycle of periodicity would be most useful to acquire, with relatively smaller gains thereafter.

We next make a forecast for the scenario in which the fiducial Doppler boost model is incorrect, either because the parameters we adopted differ from the true values, or because Doppler modulations are entirely absent from the true variability of PG1302-102. We generated  mock data to examine examples for both of these cases. In the first case, we assumed that the true Doppler boost ratio is higher than in our fiducial model (e.g., $A_{DB}=3$ instead of 2.17). In the second case, we assumed that PG1302-102's variability is caused by DRW alone. The middle and right panel of Figure~\ref{Fig:example_LCs} demonstrate examples of each case, along with the fiducial model. In both cases, we expect that as new data is added, the $p$-values would begin to decrease, since the wrong model is being fit to more and more data. Our results, shown by the light blue and dark red curves in Figure~\ref{Fig:forecast}, indeed show these trends in the long run.  However, while in the latter case (when the true variability is a pure DRW), the $p$-values decrease monotonically, the former case, in which the Doppler amplitude ratio is guessed incorrectly, shows a temporary increase over the first additional cycle. These results lead us to conclude that converging on the correct model will require monitoring PG1302-102 for at least two additional cycles of the periodicity. On the other hand, just $\approx$ two additional years of data appears very useful to distinguish between the pure DRW and the DRW+DB hypotheses.

\vspace{\baselineskip}

\subsection{Future Work}
\label{sec:limitations}
We have described several caveats and limitations of the currently available data that prevent us from definitively ruling out the multi-band observations of PG1302-102 are consistent with the Doppler boost model. Here we delineate potential improvements in the observing strategy that will allow us to tackle each of the above issues. 

First, we saw that we reached different conclusions from the analysis of $V$ and $B$ band observations; in \S~\ref{sec:OpticalAmplitude}, we discussed the possibility that our results are affected by the small amplitude of the {\it Swift} $V$-band light curve, e.g., compared to ASAS-SN. This may be caused by unfortunate sampling, given that we examine a small number of simultaneous optical/UV observations. If the light curve were to include a larger number of observations (e.g., twice as many points), the light curve, and thus our results, would be less prone to such effects. 

Another possibility/limitation in the fiducial values taken in this study is the assumption that the relative amplitudes are fixed and equal to 2.17. As discussed above, if the spectral indices change over time, the relative amplitude will reflect this change. If the photometric observations are accompanied by spectral measurements, practically, there will be no free parameters in the model. With simultaneous multi-band photometric and spectroscopic observations,  even a small number of data points can provide more definitive conclusions.

Finally, we incorporated the stochastic variability of quasars by adding the DRW variability of optical and UV wavelengths. We assume that these deviations are incoherent (we draw independent realisations for the optical and UV DRW variability). Even though this is beyond the scope of this paper, a comprehensive analysis of the covariance between optical and UV light curves of quasars is necessary to validate this choice (see \citealt{Xin+2020}).

\section{Summary}
\label{sec:summary}
In this paper, we presented simultaneous observations of PG1302-102 with the {\it Swift} satellite in two UV and two optical bands. This is a significant enhancement to the previously available observations, which consisted of a smaller number of UV data points that were not taken simultaneously with optical data and only in one optical and one UV band. We performed a statistical analysis to test the Doppler boost hypothesis, which predicts that the UV variability should track the optical, but with a $\sim2.2$ times higher amplitude. From the analysis of nine simultaneous observations from {\it Swift}, we found that:
\begin{itemize}
    \item The new light curves roughly trace the sinusoidal trends expected from the Doppler boost model.
    \item The multi-wavelength data are consistent with relativistic Doppler boost when the $B$-band versus $M2$ (and $W1$) data are considered. 
    \item The $V$-band versus $M2$ (and $W1$) data could still be consistent with the Doppler boost model, but only if either the ratio of UV/optical variability is larger than expected from the spectral slopes, the stochastic variability makes large contribution in the UV, or the UV/optical spectral slopes vary.
    \item A potential explanation for the rejection of the Doppler boost model (with the $V$-band data) is that the sparse new optical data from {\it Swift} underestimate the true optical variability. Comparison with the light curve from ASAS-SN suggests that this is likely. 
    \item If we consider all four bands  simultaneously, combining independent pairs of bands, the Doppler model is disfavored.
    \item Additional, simultaneous optical and UV observations tracking another cycle of PG1302-102's proposed period should lead to a definitive conclusion.
\end{itemize}

\section*{Acknowledgements}
We thank Michele Vallisneri for useful suggestions and Tingting Liu for providing the best-fitting parameters of their models. 
M.~Charisi acknowledges support from
the National Science Foundation (NSF) NANOGrav Physics
Frontier Center, award number 1430284. 
Z.~Haiman acknowledges support from NASA grants NNX17AL82G and 80NSSC19K0149 and NSF grant 1715661. The work of D.~Stern was carried out at the Jet Propulsion Laboratory, California Institute of Technology, under a contract with NASA. We acknowledge the use of public data from the {\it Swift} data archive.

\appendix
\section{{\it Swift} Data} \label{sec:appendix-data}
This appendix includes the {\it Swift} observational data for all 6 UVOT bands, along with their errors.

\begin{table*}
\centering
    \begin{tabular}{l|l|l|l|l|l|l|l|l|l|l|l|l}
    \multicolumn{1}{|c|}{MJD}& \multicolumn{2}{c|}{V} & \multicolumn{2}{c|}{B} & \multicolumn{2}{c|}{U} & \multicolumn{2}{c|}{W1} & \multicolumn{2}{c|}{M2} & \multicolumn{2}{c|}{W2} \\ \hline
        56892& 15.26 &0.07  & 15.38 & 0.03 &14.17 & 0.03 & 14.10&  0.03 & 14.08 & 0.01 & 14.12 & 0.02   \\
        57456& 15.28 & 0.05  & 15.27 & 0.03 &14.25 & 0.03 & 14.14& 0.03& 14.24 & 0.01  &  14.17 & 0.02 \\ 
        57859& 15.29& 0.04  & 15.52&  0.03 & 14.23 & 0.02 & 14.22 & 0.01 & 14.10 & 0.01 & 14.27 & 0.02 \\ 
        57972& 15.29 & 0.04 &15.42&  0.03 & 14.21& 0.01 & 13.88 & 0.02  &  13.91& 0.02 & 13.90 & 0.02\\ 
        58086& 15.35 &0.03 & 15.57 &0.02 &14.41 & 0.02& 14.20& 0.02 & 14.14 & 0.01 & 14.28 & 0.02 \\ 
        58192& 15.29 &0.03 &15.45 &0.02 & 14.20 & 0.01& 14.01 & 0.02 & 14.03 & 0.02 & 14.00 & 0.02 \\ 
        58269& 15.28 &0.02& 15.37& 0.02& 14.17& 0.01& 13.89 & 0.01 & 13.83 & 0.01 & 13.87 & 0.01 \\ 
        58337&15.23 &0.03 & 15.31 & 0.01 & 14.11 & 0.01 & 13.79 & 0.01 & 13.55 & 0.01 & 13.72 & 0.01 \\ 
        58546&15.06&0.02 & 15.12 & 0.01 & 13.90& 0.01 & 13.37 & 0.01 & 13.36 & 0.01 & 13.26 & 0.01 \\ \hline
    \end{tabular}
    \caption{{\it Swift} Data. Column 1: MJD; Column 2,4,6,8,10,12: $V$, $B$, $W1$, $M2$, $W2$-band magnitudes; Column 3,5,7,9,11,13: $V$, $B$, $W1$, $M2$, $W2$-band magnitude errors.}
    \label{Table:swift_data}

\end{table*}

\bibliographystyle{mnras}
\bibliography{PG1302} 
\bsp
\label{lastpage}
\end{document}